\documentclass[prb,aps,twocolumn,eqsecnum,showpacs]{revtex4}
\usepackage{graphicx}

\usepackage{epsfig}
\usepackage{graphicx}
\usepackage{dcolumn}
\usepackage{bm}

\begin{document}

\title{       Spin-orbital resonating valence-bond liquid on a triangular lattice:\\
              Evidence from finite cluster diagonalization }

\author{      Ji\v{r}\'i Chaloupka }
\affiliation{ Max-Planck-Institut f\"ur Festk\"orperforschung,
              Heisenbergstrasse 1, D-70569 Stuttgart, Germany }
\affiliation{ Department of Condensed Matter Physics, Faculty of Science,\\
              Masaryk University, Kotl\'a\v{r}sk\'a 2, CZ-61137 Brno,
              Czech Republic }

\author{      Andrzej M. Ole\'s }
\affiliation{ Max-Planck-Institut f\"ur Festk\"orperforschung,
              Heisenbergstrasse 1, D-70569 Stuttgart, Germany }
\affiliation{ Marian Smoluchowski Institute of Physics,
              Jagellonian University, Reymonta 4, PL-30059 Krak\'ow, Poland }

\date{16 December 2010}

\begin{abstract}

We investigate the ground state of the $d^1$ spin-orbital model for
triply degenerate $t_{2g}$ orbitals on a triangular lattice which
unifies intrinsic frustration of spin and orbital interactions with
geometrical frustration. Using full or Lanczos exact diagonalization of
finite clusters we establish that the ground state of the spin-orbital
model which interpolates between the superexchange and direct exchange
interactions on the bonds is characterized by valence-bond correlations.
In the absence of Hund's exchange the model describes
a competition between various possible valence-bond states.
By considering the clusters with open boundary conditions we demonstrate
that orbital interactions are always frustrated, but this frustration is
removed by pronounced spin singlet correlations which coexist with
supporting them dimer orbital correlations. Such local configurations
contribute to the disordered ground states found for the clusters with
periodic boundary conditions which interpolate between a highly
resonating, dimer-based, entangled spin-orbital liquid phase, and
a valence-bond state with completely static spin-singlet states. We
argue that these states are also realized for the infinite lattice and
anticipate that pronounced transitions between different regimes found
for particular geometries will turn out to smooth crossovers in the
properties of the spin-orbital liquid in the thermodynamic limit.
Finally, we provide evidence that the resonating spin-orbital liquid
phase involves entangled states on the bonds. In such a phase classical
considerations based on the mean-field theory cannot be used, spin
exchange interactions do not determine spin bond correlations, and
quantum fluctuations play a crucial role in the ground states and
magnetic transitions.\\
{\it Published in: Physical Review B \textbf{83}, 094406 (2011).}

\end{abstract}


\maketitle

\section{Spin-orbital frustration}

Frustration in magnetic systems is usually of geometrical origin,
but it may also arise due to competing exchange interactions.
\cite{Diep,Hfm,Faz99,Nor09,Bal10} A common feature of frustrated spin
systems is that the interactions along different bonds compete with one
another, and this leads in some cases to disordered states and to
quantum phase transitions when the interaction strength is
varied. Another possibility is so-called "order-by-disorder",
and several microscopic mechanisms which stabilize ordered state in
spin systems have been investigated.\cite{Hfm,Faz99,Nor09,Bal10}
Here we shall focus on the triangular lattice, where frustrated
interactions suggest that valence-bond configurations in spin model
with antiferromagnetic (AF) interactions could play an important role,
\cite{Faz74} and we supplement them in this work by the frustrated
orbital degrees of freedom.

Recently interesting physical realizations of frustrated interactions
were introduced in the context of spin-orbital superexchange which
arises in transition metal oxides with active orbital degrees of freedom.
\cite{Kug82,Ole09} In such models frustration is intrinsic and
follows from the directional nature of orbital interactions.\cite{Fei97}
Therefore, the orbital part of the spin-orbital superexchange is
frustrated even {\it without\/} any geometrical frustration. Generic
features of this direction-dependent interactions are captured within
the two-dimensional (2D) quantum compass model,\cite{Nus04} which
exhibits a quantum phase transition through an isotropic point with
highly degenerate ground state (GS).\cite{Dor05,Wen08,Oru09}
This high degeneracy is a fingerprint of highly frustrated interactions
and occurs also in the one-dimensional (1D) compass model.\cite{Brz07}
It is also characterized by a rather surprising hidden dimer order in
the GS which follows from the symmetry of compass interactions.\cite{Brz10}

Frustration in the orbital superexchange models is somewhat more subtle
--- the interactions depend on the type of active orbital degree
of freedom and in each case differ from those in the quantum compass
model. In case of $e_g$ orbitals the interactions are directional as in
the quantum compass model, but they are Ising-type only for one cubic
axis, e.g. for the bonds $\langle ij\rangle$ along the $c$ cubic axis
one has the interaction $\propto\sigma_i^z\sigma_j^z$, while for the
bonds $\langle ij\rangle$ in the $ab$ planes they involve linear
combinations of $\{\sigma_i^z,\sigma_i^x\}$ operators which arise from
the directional orbital states along the considered ($a$ or $b$) axis
\cite{Fei97,Fei99} (here $\sigma_i^z$ and $\sigma_i^x$ are Pauli
matrices). This particular structure follows from the fact that
although only one of $e_g$ orbital states participates in charge
excitations along each single cubic axis and the interactions
appear to be classical in a 1D $e_g$ orbital model,\cite{Dag04}
their superposition is quantum either in a 2D model,\cite{Dag06}
or in a three-dimensional one.\cite{Fei05}
In contrast, two $t_{2g}$ orbitals are active and participate in
charge excitations along each cubic direction, so the respective
interactions involve {\it a priori\/} all three components of the
orbital pseudospin $\tau=1/2$ doublet, with the restriction that
the active orbital $t_{2g}$ doublet changes with the cubic axis.
\cite{Kha00} For instance, where the degeneracy of $t_{2g}$ orbitals is
removed by crystal field in the vanadium perovskites and $xy$ orbitals
are filled, the $\{yz,zx\}$ orbital doublet contributes with orbital
fluctuations to the bonds along the $c$ axis.\cite{Kha01,Hor08}

Realistic superexchange models for perovskite transition metal oxides
include both orbital and spin degrees of freedom, which are strongly
interrelated.\cite{Kug82,Ole05} Two important questions for these models
are: (i) whether the orbital frustration can be removed by properly
selected spin states, or frustration is even enhanced by spin-orbital
quantum fluctuations, and (ii) to what extent spin dynamics may be
treated as independent of orbital dynamics.\cite{Ole06} The disordered
GS was suggested for the $t_{2g}$ orbitals in $d^1$
configuration on the perovskite lattice.\cite{Kha00}

In the present paper we want to focus on the model derived for the
transition metal ions in $d^1$ configuration for the triangular
lattice,\cite{Nor08} with frustration
being both of orbital and geometrical origin. This spin-orbital model
corresponds to the undistorted NaTiO$_2$ and describes magnetic
interactions in the spin-orbital space, with superexchange and direct
exchange. In the direct exchange case the model is exactly solvable and
the GS was determined by considering the dimer coverings of the lattice,
with each dimer containing a spin singlet accompanied by two active
orbitals on the direct exchange bond.\cite{Jac07} In a general case the
GS and the ratio of superexchange and direct exchange are not known ---
the latter depends on the respective effective hopping elements via the
oxygen orbitals responsible for the superexchange and the $(dd\sigma)$
hopping which gives the direct exchange. Therefore, we use it below as
a model parameter. A second parameter of the spin-orbital model
considered here is Hund's exchange interaction. One might expect that
also in case of superexchange interactions $t_{2g}$ orbitals could
order and remove the frustration in the triangular lattice, as they do,
for instance, in LiVO$_2$.\cite{Pen97} It was argued, however, in Ref.
\onlinecite{Nor08} that the GS of the present $d^1$ spin-orbital model
is disordered and dominated by dimer correlations practically for any
ratio of the superexchange and direct exchange interaction.
This conclusion was drawn by considering the mean-field (MF) states,
variational wave functions with valence-bond correlations, and exact
diagonalization of small systems of not more than $N=4$ sites.

The purpose of this paper is to reanalyze the spin-orbital states in
the $d^1$ spin-orbital model on the triangular lattice,\cite{Nor08}
and provide evidence in favor of the disordered spin-orbital liquid GS
from numerical studies of larger finite systems, having up to
$N=10$ sites. We use extensively Lanczos diagonalization, but for
rather small systems of size up to $N=6$ sites where full
diagonalization is also possible in the subspace of ${\cal S}^z=0$
total spin, both methods were compared with each other. Thereby,
we addressed a few general questions which concern the spin-orbital
physics for varying parameters of the model:
(i) nature of dimer spin and orbital correlations,
(ii) nature of the transition to the spin-polarized ferromagnetic
(FM) state with increasing Hund's exchange,
(iii) importance of spin-orbital entanglement\cite{Ole06} and its
consequences for the transition from low-spin to high-spin states.
We will provide answers to these questions by considering systems
of various size and with different boundary conditions. Altogether,
we shall demonstrate that quantum fluctuations determine the GS and the
magnetic transitions to such an extent that classical considerations
cannot be used in several situations.

The paper is organized as follows. In Sec.~\ref{sec:sex} we introduce
the $d^1$ spin-orbital model on a triangular lattice (as for Ti$^{3+}$
or V$^{4+}$ ions) as derived in Ref. \onlinecite{Nor08}.
Basic information about the cluster sizes and geometries used in
Lanczos diagonalization is contained in Sec. \ref{sec:lan}.
The numerical results obtained for the isolated clusters of
up to $N=10$ sites are analyzed in Sec. \ref{sec:sin}. In this section
we investigate the model in absence of Hund's exchange and analyze
bond correlations: spin, orbital and spin-orbital ones, as well as the
orbital occupation. They allowed us to find
certain general trends which are expected to determine the
behavior of the model in the thermodynamic limit. The generic
transition from quantum to classical regime in the singlet sector,
with interactions evolving from superexchange to direct exchange, is
illustrated by a hexagonal cluster in Sec. \ref{sec:hex}. Next we
consider triangular clusters with open boundary conditions in Sec.
\ref{sec:tri} and show that the singlet correlations are robust in
the entire regime of the exchange interactions. The results obtained
for the clusters with periodic boundary conditions are reported
in Sec. \ref{sec:pbc}. In Sec. \ref{sec:fm} we present the orbital
model obtained in spin polarized case, and investigate the
transition from low-spin to high-spin states for a few representative
clusters, presenting the respective phase diagrams in Sec.
\ref{sec:phd}. Finally, we present the consequences of spin-orbital
entanglement in Sec. \ref{sec:dis} and show that it modifies the phase
diagrams significantly with respect to those obtained when spin and
orbital operators are disentangled, particularly in the regime of
purely superexchange interactions. We also point out in Sec.
\ref{sec:exc} that meaningful exchange constants cannot be introduced
in cases where spin-orbital entanglement dominates and stabilizes the
low-spin ground state with large spin-orbital fluctuations.
General discussion and summary are presented in Sec.~\ref{sec:summa}.

\section{Spin--orbital model }
\label{sec:som}

\subsection{Superexchange versus direct exchange}
\label{sec:sex}

We consider the spin-orbital model on the triangular lattice derived
in Ref. \onlinecite{Nor08} which describes interactions between
$S=1/2$ spins for $d^1$ electron configurations, such as in NaTiO$_2$.
The magnetic transition metal ions form a triangular lattice for the
$\langle 111\rangle$ planes of a compound with cubic symmetry. The
bonds $\langle ij\rangle$ are spanning the three directions, labeled by
$\gamma = a,b,c$.

\begin{figure}[t!]
\begin{center}
\includegraphics[width=8.2cm]{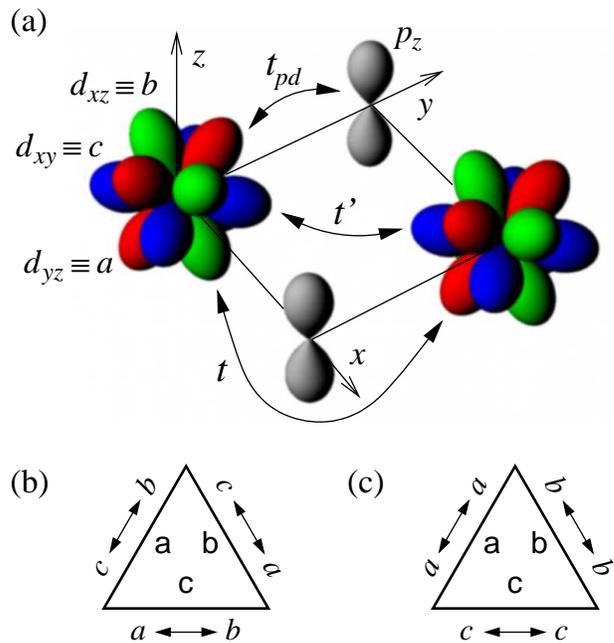}
\end{center}
\caption{(Color online) (a) Schematic view of the hopping
processes between $t_{2g}$ orbitals along a bond parallel to the $c$
axis in NaTiO$_2$:
(i) $t_{pd}$ between Ti($t_{2g}$) orbitals and O($2p_z$) orbitals, with
two $t_{pd}$ transitions contributing to an effective hopping $t$, and
(ii) direct $d-d$ hopping $t'$.
The hopping $t$ interchanges two orbital flavors on two sites and
contributes to the effective superexchange interactions on a bond in
the triangular lattice, while the latter (diagonal) hopping element
$t'$ contributes to the direct (kinetic) exchange. The $t_{2g}$
orbitals are shown by different gray scale (color) and are labeled as
$a$, $b$, and $c$, see Eq. (\ref{abc}).
In the bottom part the hopping processes contributing to
(b) superexchange and
(c) direct exchange, are shown for the bonds along $\gamma=a,b,c$
axes in the triangular lattice.}
\label{fig:hops}
\end{figure}

In order to explain the physical content of the model
we consider a representative bond along the $c$ axis shown in Fig.
\ref{fig:hops}(a). For the realistic parameters of NaTiO$_2$ the $3d$
electrons are almost localized in $d^1$ configurations of Ti$^{3+}$
ions, hence their interactions with neighboring sites can be described
by the effective superexchange and kinetic exchange processes. Virtual
charge excitations, $d_i^1d_j^1\rightleftharpoons d_i^2d_j^0$,
between the neighboring sites generate magnetic interactions which
arise from two different hopping processes for active $t_{2g}$ orbitals:
(i) the effective hopping $t=t_{pd}^2/\Delta$ which occurs via oxygen
$2p_z$ orbitals with the charge transfer excitation energy $\Delta$
and consists of two $t_{pd}$ steps,\cite{Zaa93} in the present case
along the 90$^{\circ}$ bonds, and
(ii) direct hopping $t'$ which couples the $t_{2g}$ orbitals along the
bond and give direct (kinetic exchange) interaction.
Note that the latter processes couple orbitals with the same flavor,
while the former ones couple different orbitals, and there the occupied
orbitals may be interchanged as a result of a virtual charge excitation.

For convenience, we introduce the notation
\begin{equation}
\label{abc}
|a \rangle \equiv |yz\rangle\,, \hskip .7cm
|b \rangle \equiv |xz\rangle\,, \hskip .7cm
|c \rangle \equiv |xy\rangle\,,
\end{equation}
for the three $t_{2g}$ orbital flavors (colors),
following the one used in the perovskite systems,\cite{Kha00}
adopted here for the triangular lattice.\cite{Kha04}
It follows from the symmetry of orbital wave functions that only two
of the three $t_{2g}$ orbitals allow for $d-p$ hopping $t_{pd}$ and
are active in superexchange on any given bond $\langle ij\rangle$ [Fig.
\ref{fig:hops}(b)], while the remaining $\gamma$ orbitals couple
directly along the $\gamma$ axis, so they contribute to the direct
(kinetic) exchange, see Fig. \ref{fig:hops}(c). In addition, each site
is occupied by precisely one electron, so the density operators satisfy
a local constraint at each site $i$,
\begin{equation}
\label{const}
n_{ia}+n_{ib}+n_{ic}\equiv 1\,.
\end{equation}
These symmetry properties on the triangular lattice are analogous to
those which decide about the form of the kinetic energy for $t_{2g}$
electrons in the perovskite lattice.\cite{Kha00,Kha01}

Local Coulomb interactions at transition metal ions are described by
two parameters: intraorbital Coulomb interaction $U$ and Hund's
exchange $J_H$.\cite{Ole83} In the limit of large intraorbital Coulomb
interaction $U$ intersite charge excitations are transformed away and
one finds the following Hamiltonian,\cite{Nor08}
\begin{equation}
\label{som} {\cal H} = J \left\{ (1 - \alpha) \; {\cal H}_s
                 + \sqrt{(1 - \alpha) \alpha} \; {\cal H}_m
                 + \alpha \; {\cal H}_d \right\}\,,
\end{equation}
where $J$ is the exchange energy. The parameter $\alpha$ is the first
parameter of the present model Eq. (\ref{som}) and is given by the
hopping elements as follows,
\begin{equation}
\label{alpha}
\alpha=\frac{t'^2}{t^2+t'^2}
\end{equation}
--- it interpolates
between the superexchange ($\alpha=0$) and direct exchange ($\alpha=1$)
limit, as explained in Ref. \onlinecite{Nor08}. The second parameter of
the spin-orbital model Eq. (\ref{som}) is Hund's exchange,
\begin{equation}
\label{eta}
\eta = \frac{J_H}{U}.
\end{equation}
It enters the superexchange and direct exchange (see below) via
the coefficients
\begin{equation}
\label{rr}
r_1 = \frac{1}{1 - 3 \eta},  \hskip 0.5cm
r_2 = \frac{1}{1 - \eta},    \hskip 0.5cm
r_3 = \frac{1}{1 + 2 \eta},
\end{equation}
which follow from the multiplet structure of $d^2$ ions.\cite{Gri71}
They correspond to the triplet excitation at energy $(U-3J_H)$, and
to singlet excitations at energies $(U-J_H)$ and $(U+2J_H)$. Although
the actual values of the Coulomb and Hund's exchange elements were
deduced from the spectroscopic data for Ti$^{2+}$ ions corresponding
to charge excitations by Zaanen and Sawatzky,
\cite{Zaa90} $U=4.35$ eV and $J_H=0.59$ eV, which gives a value of
$\eta\simeq 0.14$, we shall use $\eta$ as a parameter below in order to
investigate the transition from low-spin to high-spin states for various
cluster size and to highlight the difference in the orbital correlations
in the low and high $\eta$ regime.

The superexchange part of ${\cal H}$ can be specified as follows
\begin{eqnarray}
\label{Hs}
{\cal H}_s \! & = & \! \frac{1}{2} \sum_{\langle ij \rangle \parallel\gamma}
\left\{ r_1\Big( \vec{S}_i \! \cdot \! \vec{S}_j + \frac{3}{4} \right)
  \Big[ A_{ij}^{(\gamma)} \! + \frac{1}{2} (n_{i\gamma} + n_{j\gamma})
  - 1 \Big] \nonumber \\
 & + & \! r_2 \left( \vec{S}_i \! \cdot \! \vec{S}_j - \frac{1}{4}\right)
\Big[ A_{ij}^{(\gamma)} \! - \frac{1}{2} (n_{i\gamma} + n_{j\gamma}) + 1
\Big] \nonumber \\ & - & \! \frac{2}{3} (r_2-r_3) \left( \vec{S}_i \!
\cdot \! \vec{S}_j - \frac{1}{4} \right) B_{ij}^{(\gamma)} \Big\}\,,
\end{eqnarray}
and contains two spin operators --- a projection on the triplet state
 $(\vec{S}_i \cdot \vec{S}_j + \frac{3}{4})$, and an operator
$-(\vec{S}_i \cdot \vec{S}_j - \frac{1}{4})$ which is a projection on
the singlet state. These operators accompany the coefficients
$\{r_1,r_2,r_3\}$ and express the dependence on the excited $d^2$
states.

The orbital operators $A_{ij}$ and $B_{ij}$ in Eq.~(\ref{Hs})
depend on the bond direction $\gamma$ and involve two
orbital flavors active in the superexchange,
\begin{eqnarray}
\label{som1abn} \hskip -.7cm
A_{ij}^{(\gamma)} \!\! & = & \!\!
\Big( T_{i\gamma}^+ T_{j\gamma}^+ \!
 + T_{i\gamma}^- T_{j\gamma}^- \Big) \! - \! 2T_{i\gamma}^z T_{j\gamma}^z
 + \! \frac{1}{2} n_i^{(\gamma)} n_j^{(\gamma)}\,,\\
\label{som1ab}
\hskip -.7cm B_{ij}^{(\gamma)} \! & = & \!\! \Big( T_{i\gamma}^+
T_{j\gamma}^- \! + T_{i\gamma}^- T_{j\gamma}^+ \Big) \! - \!
2T_{i\gamma}^z T_{j\gamma}^z + \! \frac{1}{2} n_i^{(\gamma)}
n_j^{(\gamma)}\,.
\end{eqnarray}
The operator $n_{i\gamma}$ in Eq.~(\ref{Hs}) stands for the number of
electrons at site $i$ in the orbital {\it inactive\/} in superexchange
processes, for instance $n_{i\gamma}\equiv n_{ic}$ in the example
depicted in Fig. \ref{fig:hops}. On the contrary, $n_i^{(\gamma)}$ is
the total electron number operator at site $i$ for orbitals
{\it active\/} in superexchange, i.e., in the
case shown in Fig. \ref{fig:hops} it is $n_i^{(c)}=n_{ia}+n_{ib}$.
The orbital operators $\{T_{i\gamma}^+,T_{i\gamma}^-,T_{i\gamma}^z\}$
refer to the orbital doublet active in the superexchange on the bond
$\langle ij\rangle\parallel\gamma$. For a single bond, the orbital
operators in Eqs.~(\ref{som1abn}) and (\ref{som1ab}) may be written
in a very suggestive form by performing a local transformation in which
the active orbitals are exchanged on one bond site,\cite{Kha04} e.g.
$|a \rangle \rightarrow |b \rangle$ and $|b \rangle \rightarrow |a
\rangle$ on bond $\langle ij\rangle\parallel c$:
\begin{eqnarray}
\label{som2abn}
\hskip -.7cm
A_{ij}^{(\gamma)}  & \equiv &  2 \left\{\left( {\vec T}_{i} \cdot {\vec
T}_{j}\right)^{(\gamma)}
+  \frac{1}{4}\, n_i^{(\gamma)} n_j^{(\gamma)} \right\}\,,\\
\hskip -.7cm
B_{ij}^{(\gamma)}  & \equiv &  2 \left\{\left( \vec{T}_{i} \odot
\vec{T}_{j}\right)^{(\gamma)}
+ \frac{1}{4}\, n_i^{(\gamma)} n_j^{(\gamma)} \right\}\,.
\end{eqnarray}
Here the scalar product in $A_{ij}$ is the conventional expression
for pseudospin $T=1/2$ variables transformed as described above, and
the product in $B_{ij}$ is the usual term which follows from
the structure of local Coulomb interactions, as well transformed ---
they are defined as follows:
\begin{eqnarray}
\label{ttscalar}
\left(\vec{T}_{i} \cdot \vec{T}_{j}\right)^{(\gamma)}\! &\equiv&
\frac12\Big( T_{i\gamma}^+T_{j\gamma}^- + T_{i\gamma}^-T_{j\gamma}^+\Big)
+ T_{i\gamma}^z T_{j\gamma}^z\,,  \\
\label{ttcross}
\left(\vec{T}_{i} \odot \vec{T}_{j}\right)^{(\gamma)}\! &\equiv&
\frac12\Big(T_{i\gamma}^+T_{j\gamma}^+ + T_{i\gamma}^-T_{j\gamma}^-\Big)
+ T_{i\gamma}^z T_{j\gamma}^z\,.
\end{eqnarray}
This form follows from the local transformation at site $j$ which is
introduced for the superexchange in the present case.\cite{Kha04}
These operators select favored orbital configurations on two
neighboring sites via the $T_{i\gamma}^zT_{j\gamma}^z$ terms, and
orbital fluctuations are described by the
$T_{i\gamma}^\pm T_{j\gamma}^\mp$ and $T_{i\gamma}^\pm T_{j\gamma}^\pm$
terms. Note that the $z$-th pseudospin component is not conserved.
For a bond along the axis $\gamma$ orbital operators at site $i$ are
defined by the electron creation
$\{a_i^\dagger,b_i^\dagger,c_i^\dagger\}$ and annihilation
$\{a_i,b_i,c_i\}$ operators for fermions with a given flavor.
For instance, for the bonds along the $a$ or $b$ axis they are:
\begin{eqnarray}
\label{TaTb}
T_{ia}^+ \!& =&\! b_i^{\dagger} c_i^{}\,, \hskip 1.7cm
T_{ib}^+   =      c_i^{\dagger} a_i^{}\,,
\nonumber \\
T_{ia}^- \!& =&\! c_i^{\dagger} b_i^{}\,, \hskip 1.7cm
T_{ib}^-   =      a_i^{\dagger} c_i^{}\,,
\\
\label{TzTz}
T_{ia}^z\! & =&\!\frac12 ( n_{ib} - n_{ic} )\,, \hskip .4cm
T_{ib}^z     =   \frac12 ( n_{ic} - n_{ia} )\,.
\end{eqnarray}

The direct (kinetic) exchange term involves only virtual excitations
of $\gamma$ orbitals on a bond $\langle ij \rangle \parallel\gamma$,
\begin{eqnarray}
\label{Hd}
{\cal H}_d \! & = & \! \frac{1}{4} \sum_{\langle ij \rangle \parallel \gamma}
\Big\{ \left[ -r_1 \left( \vec{S}_i \! \cdot \! \vec{S}_j + \frac{3}{4}\right)
 + r_2 \left( \vec{S}_i \! \cdot \! \vec{S}_j - \frac{1}{4} \right) \right]
\nonumber \\ & & \hskip 1cm \times \Big[ n_{i\gamma} (1 - n_{j\gamma})
 + (1 - n_{i\gamma}) n_{j\gamma} \Big] \nonumber \\ & + & \frac{1}{3}
\left( 2r_2 + r_3 \right) \left( \vec{S}_i \! \cdot \! \vec{S}_j - \frac{1}{4}
\right) \; 4 n_{i\gamma} n_{j\gamma} \Big\}\,.
\end{eqnarray}
Therefore the structure of the orbital operators is here simpler ---
they enter as projection operators and give two different individual
terms:\cite{Nor08} either (i) when only one active orbital is occupied,
$\propto n_{i\gamma}(1-n_{j\gamma})$, or (ii) when both orbitals are
occupied, $\propto n_{i\gamma}n_{j\gamma}$. For the bond shown in
Fig. \ref{fig:hops} $\gamma\equiv c$. The structure of $d^2$ excited
states is here the same as for the superexchange, so the same
coefficients given in Eqs. (\ref{rr}) occur in both terms,
superexchange Eq. (\ref{Hs}) and kinetic exchange Eq. (\ref{Hd}).

As explained in Ref. \onlinecite{Nor08}, the two different types of
hopping processes ($t$ and $t'$) may contribute in a two-step virtual
$d_i^1 d_j^1 \rightleftharpoons d_i^2d_j^0$ excitation, in such a way
that the occupied orbitals are changed at both sites. In such a case
the resulting effective interaction are expressed in terms of orbital
fluctuation operators. For the bond shown in Fig. \ref{fig:hops} these
terms are:
\begin{eqnarray}
\label{Hm}
{\cal H}_m^{(c)} \! & = & \! - \frac{1}{4} \sum_{\langle ij \rangle
\parallel c} \left\{  r_1 \left( \vec{S}_i \! \cdot \! \vec{S}_j
 + \frac{3}{4} \right) - r_2 \left( \vec{S}_i \! \cdot \! \vec{S}_j
 - \frac{1}{4} \right) \right\} \nonumber \\ & \times & \! \Big( T_{ia}^+
T_{jb}^+ + T_{ib}^- T_{ja}^- + T_{ib}^+ T_{ja}^+ + T_{ia}^- T_{jb}^- \Big)\,,
\end{eqnarray}
where the orbital operators are defined in Eqs. (\ref{TaTb}).
The form of the ${\cal H}_m^{(a)}$ and ${\cal H}_m^{(b)}$ terms is
obtained from Eq. (\ref{Hm}) by cyclic permutations of the orbital
indices. Note that these terms describe fluctuations that go
{\it beyond\/} any static orbital configuration, so they
represent corrections to the classical treatment of the spin-orbital
correlations, as discussed in Secs. \ref{sec:dis} and \ref{sec:exc}.

In the subsequent sections we will focus first on the frustrated
interactions in the model of Eq.~(\ref{som}) at $\eta = 0$. This case
is rather special as the multiplet structure collapses to a single
excitation with energy $U$ (spin singlet and triplet excitations are
then degenerate), and the Hamiltonian depends only on the ratio of
superexchange to direct exchange, parametrized by $0 \le \alpha \le 1$,
and has the following form:
\begin{eqnarray}
\label{som0}
{\cal H}_0\!\! &=&\! \!
J \sum_{\langle ij \rangle \parallel \gamma}
\Big\{ (1 - \alpha) \left[ 2 \left( \vec{S}_i \cdot \vec{S}_j
 + \frac{1}{4} \right)\right. \nonumber \\
& &\!\!\left.\times\left[
\left( \vec{T}_{i} \cdot \vec{T}_{j}\right)^{(\gamma)}
 + \frac{1}{4} n_i^{(\gamma)} n_j^{(\gamma)} \right]\!
 + \frac{1}{2} (n_{i\gamma} + n_{j\gamma}) - 1 \right] \nonumber \\
& +&\!\!  \alpha \left[ \left( \vec{S}_i \! \cdot \! \vec{S}_j
 - \frac{1}{4} \right) n_{i\gamma} n_{j\gamma}\right.  \nonumber \\
& & \!\left. \hskip .3cm
-\frac{1}{4} \Big(
n_{i\gamma} (1 - n_{j\gamma}) + (1 - n_{i\gamma}) n_{j\gamma} \Big)
\right] \nonumber \\
&- &\! \!\frac{1}{4} \sqrt{\alpha (1 - \alpha)} \; \Big(
T_{i\bar{\gamma}}^+ T_{j{\tilde \gamma}}^+ + T_{i{\tilde \gamma}}^-
T_{j\bar{\gamma}}^- + T_{i{\tilde \gamma}}^+ T_{j\bar{\gamma}}^+
+ T_{i\bar{\gamma}}^- T_{j{\tilde \gamma}}^- \Big)
\Big\}.\nonumber \\
\end{eqnarray}
Here the orbital scalar product
$\left( \vec{T}_{i} \cdot \vec{T}_{j}\right)^{(\gamma)}$ is given by Eq.
(\ref{ttscalar}). The Hamiltonian ${\cal H}_0$ describes AF interactions
between spins, but the orbital terms favor either orbital fluctuations
(superexchange) or a static configuration of the same orbitals (direct
exchange) on the bond.
The form of the superexchange $\propto (1-\alpha)$ suggests that the
spin and orbital sectors could be completely equivalent and symmetrical
at $\alpha = 0$ for a single bond. However, this remains true only as
long as active orbitals can be selected to contribute to the
superexchange as this bond, i.e., for $n_i^{(\gamma)}n_j^{(\gamma)}\equiv 1$,
and this equivalence is broken when more bonds are considered.\cite{Nor08}

A remarkable feature of the Hamiltonian Eq. (\ref{som0}) is the lack of
higher symmetry in any of the points when $\alpha$ is varied. Even at
$\alpha=0.5$, where all electron transitions have the same amplitude,
no higher symmetry occurs as the superexchange ($\alpha=0$) and direct
exchange $\alpha$ result from quite distinct processes and cannot be
transformed into each other. The only analytical solution was found in
$\alpha=1$ case, where at $\eta=0$ the extremely degenerate GS is a
liquid of hard-core dimers.\cite{Jac07} This degeneracy is removed at
$\eta>0$, and a valence-bond crystal with a large unit cell of 20 sites
is formed.

\subsection{Lanczos diagonalization and cluster size}
\label{sec:lan}

In order to establish unbiased results concerning the nature of the GS
and spin and orbital correlations in the present spin-orbital model Eq.
(\ref{som}), providing evidence in favor of spin-orbital liquid state,
we have used diagonalization of finite
clusters. As the number of degrees of freedom per site is $2\times3=6$,
the size of the Hilbert space increases very fast with increasing
system size $N$. The only symmetry which could straightforwardly be
implemented is the conservation of the $z$-th component of total spin,
\begin{equation}
\label{Sz}
{\cal S}^z=\sum_{i=1}^NS_i^z\,,
\end{equation}
while the orbital state is
{\it a priori\/} undetermined. Therefore, the use of full exact
diagonalization is practically limited to system of up to $N=6$ sites,
where the size of the Hilbert space is $6^6=46656$ and the largest
(${\cal S}^z=0$) subspace has the dimension $14580$. The use of the
Lanczos method allowed us to investigate systems of up to $N=10$ sites.

\begin{figure}[t!]
\includegraphics[width=7.5cm]{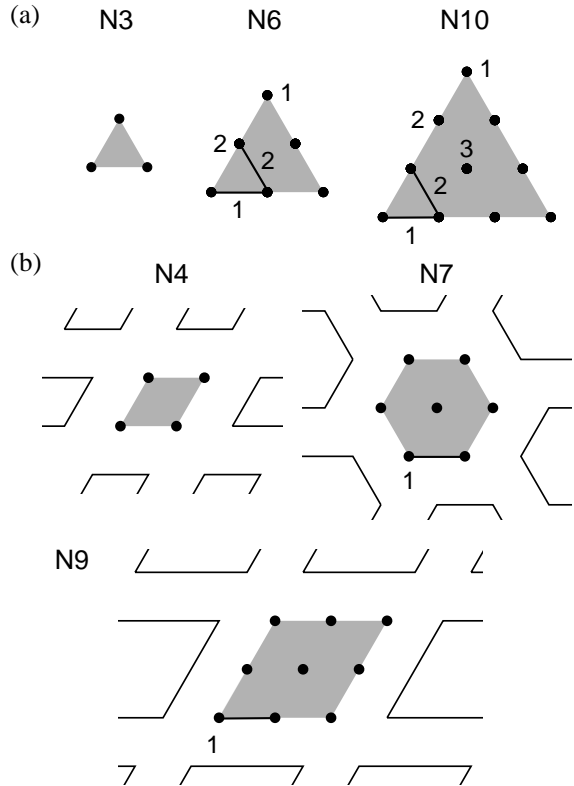}
\caption{Clusters (shown in gray) used to investigate the
spin-orbital model (\ref{som}) by full or Lanczos exact
diagonalization:
(a) triangular clusters with open boundary
conditions N3, N6 and N10 --- here nonequivalent sites and bonds used
later are indicated by labels 1, 2 and 3, and
(b) clusters with periodic boundary conditions: rhombic N4, hexagonal
N7 and rhombic N9. In the latter case positions of neighboring clusters
which cover the triangular lattice are indicated and all the sites and
bonds are equivalent (label 1 is used for equivalent sites and bonds in
N7 and N9 clusters). }
\label{fig:clu}
\end{figure}

Although several other clusters were studied as well, we would like to
concentrate here only on two classes of clusters which help to identify
certain general trends for the present spin-orbital model. First of
them are triangular clusters with open boundary conditions (OBC) which
by construction have nonequivalent sites and are expected to favor dimer
correlations. They were used to identify the dimer correlations and to
find their evolution with increasing size. These clusters help to
understand the interrelation between spin and orbital states on the
bonds which become transparent when the symmetry in the orbital space
is broken by geometry. The clusters considered here contain $N=3$, 6, 10
sites and are labeled N3, N6 and N10, see Fig. \ref{fig:clu}(a).

The second class of clusters, shown in Fig. \ref{fig:clu}(b), consists
of three clusters with up to 9 sites which cover entirely the
triangular lattice and can thus be investigated using periodic boundary
conditions (PBC): rhombic cluster N4, hexagonal cluster N7, and large
rhombic cluster N9. However, the PBC are not unique and, as we have
verified, lead to nonequivalent results. Therefore we selected in each
case displayed in Fig. \ref{fig:clu}(b) such PBC which result from
ordering the considered clusters in
rows on the triangular lattice, as indicated by the clusters surrounding
the one used for Lanczos (or full) diagonalization.
Unlike some other lattice coverings, these boundary conditions
guarantee that all the bonds and sites are equivalent in each considered
cluster. Therefore, no additional frustration of interactions
is introduced by selecting the PBC and the internal symmetry of the
considered cluster is preserved. Hence, we suggest that these clusters
may serve to simulate the situation in the thermodynamic limit.

\subsection{Correlation functions and entanglement}
\label{sec:corr}

In Secs. \ref{sec:sin}-\ref{sec:his} we will compute and discuss the
GSs of several clusters, by looking at their energies,
degeneracies, site occupations, as well as the spin, orbital and
spin-orbital (four-operator) correlation functions for a bond
$\langle ij\rangle$ along $\gamma$ axis, given respectively by
\begin{eqnarray}
\label{ss}
S_{ij}\!& \equiv & \!\frac{1}{d}\;\sum_n
\big\langle n\big|{\vec S}_i \cdot {\vec S}_j \big|n\big\rangle\,, \\
\label{tt}
T_{ij}\! & \equiv & \!\frac{1}{d}\;\sum_n\big\langle n\big|
({\vec T}_i \cdot {\vec T}_j)^{(\gamma)}\big|n\big\rangle\,,\\
\label{st}
C_{ij}\!& \equiv & \frac1d \sum_n \langle n|
(\vec S_i\cdot\vec S_j-S_{ij})(\vec T_i\cdot\vec T_j-T_{ij})^{(\gamma)} |n\rangle
\nonumber \\
\!& = & \!\frac{1}{d}\;\sum_n
\big\langle n\big| ({\vec S}_i \cdot {\vec S}_j)
     ({\vec T}_i \cdot {\vec T}_j)^{(\gamma)} \big|n\big\rangle
\\
& - & \!\frac{1}{d}
\sum_n\big\langle n\big|{\vec S}_i \! \cdot \! {\vec S}_j \big|n\big\rangle
\frac{1}{d}
\sum_m\big\langle m\big|({\vec T}_i \! \cdot \! {\vec T}_j)^{(\gamma)}
|m\big\rangle\,, \nonumber
\end{eqnarray}
where $d$ is the GS degeneracy, and the pseudospin scalar product in
Eqs. (\ref{tt}) and (\ref{st}) is defined by Eq. (\ref{ttscalar}).
In clusters with OBC the correlations
depend on the bond, but for the clusters with PBC all the bonds are
equivalent and these correlations are uniform. The summations include
all independent quantum states $\{|n\rangle\}$ which span the possibly
degenerate GS manifold.

The degeneracy $d$ of the GS
and the corresponding state vectors are easily obtained if full exact
diagonalization can be used.  On the other hand, due to appearance of
spurious degeneracies, the Lanczos diagonalization in its basic form is
not able to quantify reliably the degeneracy and to generate the set of
suitable GS vectors. In the present work we have used the
following way to remedy this problem: The Lanczos algorithm is performed
several times using random initial vectors and a sufficient number of
the GS vectors is generated. The orthonormalization of this
set then yields the degeneracy of the GS as the number of
independent vectors and the orthonormal GS vectors themselves
can be used in the GS averaging implied by Eqs. (\ref{ss})-(\ref{st}).
Furthermore, it is possible to determine the range of possible values
of the quantities of interest within the GS manifold. This is achieved
by evaluating the minimum and maximum eigenvalue of the matrix
representing the corresponding operator within this manifold. In such
cases the obtained ranges of possible values are indicated in all the
relevant figures in a form of vertical lines.

The last correlation function $C_{ij}$ Eq. (\ref{st}) quantifies the
average difference between the complete spin-orbital operator and its
decoupled product. Therefore, we use it here as the simplest measure of
spin-orbital entanglement. If $C_{ij} = 0$, the MF decoupling of spin
and orbital operators on the bond $\langle ij\rangle$ is exact and both
subsystems may be treated independently from each other. This implies
that the GS wave function can be written as a product of its
spin and orbital parts. For instance, this happens for the
high-spin states which become the GS at large $\eta$.\cite{Ole06}

However, we note that the criterion of spin-orbital entanglement
introduced above as $C_{ij}\neq 0$ can be rigorously applied only for
systems with nondegenerate GS ($d=1$). If $d>1$, the averaging
introduced in $S_{ij}$ and $T_{ij}$, used in Eq. (\ref{st}), means
that $C_{ij}$ could be (small but) finite even in cases when
spin-orbital {\it de facto\/} decoupling takes place, and the measure
of entanglement would have to be more subtle. Nevertheless, we use here
$C_{ij}$ Eq. (\ref{st}) as a simple diagnostic tool, and comment in
more detail on particular cases, where $C_{ij}\neq 0$.

\section{TOWARD SPIN-ORBITAL LIQUID}
\label{sec:sin}

\subsection{Quantum and classical dimers
in a hexagonal cluster}
\label{sec:hex}

In order to understand the consequences of frustration on spin,
orbital and spin-orbital correlations on the triangular lattice it is
instructive to start with analyzing a simpler case of a honeycomb
lattice, where the geometrical frustration is absent.\cite{Nor08}
A representative cluster for such a lattice is a hexagonal
cluster H6 (obtained from the N7 cluster in Fig. \ref{fig:clu}(b) by
removing the central site) which will be considered here with OBC.
It serves to unravel the generic behavior of the
orbital correlations for increasing $\alpha$ and to investigate the
spin-orbital entanglement in different parameter regimes of the
spin-orbital model Eq. (\ref{som}).
This cluster was analyzed using Lanczos diagonalization, and
we present also the degeneracy of the GS below.

\begin{figure}[t!]
\includegraphics[width=8.2cm]{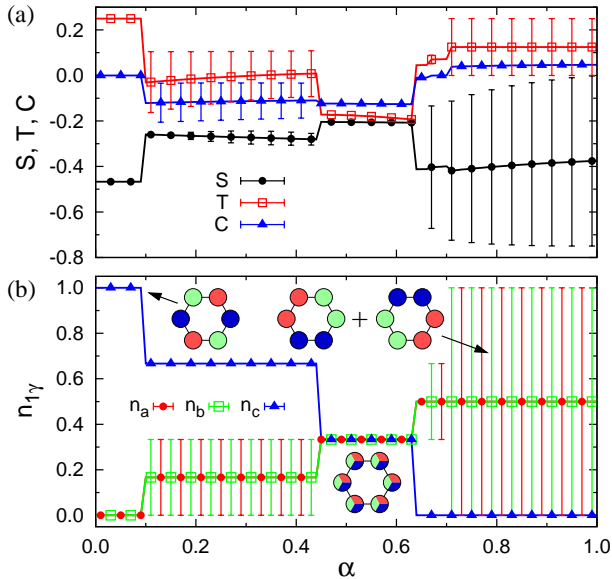}
\caption{(Color online) Evolution of the GS for a hexagon H6 with OBC
as a function of $\alpha$in the absence of Hund's exchange ($\eta=0$):
(a) bond correlations --- spin $S_{ij}$ Eq.
(\ref{ss}), circles, orbital $T_{ij}$ Eq. (\ref{tt}), squares, and
spin--orbital $C_{ij}$ Eq. (\ref{st}), triangles;
(b)~orbital electron densities
$n_{1\gamma}$ at site $i=1$ (left-most site): $n_{1a}$ (circles),
$n_{1b}$ (squares), $n_{1c}$ (triangles). The insets indicate the
orbital configurations realized in the superexchange limit ($\alpha=0$),
for $0.44<\alpha<0.63$, and in the direct exchange limit ($\alpha=1$).
The vertical lines indicate the range of possible values of the
particular quantity (bond correlation or electron density). They should
not be confused with any kind of numerical error as they indicate an
exactly determined range due to the GS degeneracy.}
\label{fig:hex}
\end{figure}

One finds that all the orbitals contribute equally in the entire range
of $\alpha$, and each orbital state is occupied at two out of six sites
in the GS, in the entire regime of $\alpha$. However, the orbital state
changes under increasing $\alpha$ and one finds four distinct regimes
by analyzing the evolution of the GS shown by the correlation functions
and orbital densities displayedin Fig. \ref{fig:hex}. The transitions
between them are abrupt (first order) and occur by level crossing.
Each site in the hexagonal cluster participates only in two bonds which
breaks the spatial symmetry in the orbital space in a particular way.
In the superexchange model at $\alpha=0$ there is precisely one orbital
at each site which contributes to the interactions along both bonds,
and we have found that indeed $n_{i\gamma}=1$ for this particular
orbital [for site $i=1$, $c$ orbital is active along both bonds,
as shown in Fig. \ref{fig:hops}(b)]. This results in a unique GS
which is characterized by a frozen orbital configuration and triplet
orbital correlations at each bond $\langle ij\rangle$, i.e.,
$T_{ij}^{\gamma}=0.25$, see Fig. \ref{fig:hex}(a). Under these
circumstances the orbitals decouple from
spins and Eq. (\ref{som0}) reduces to the AF Heisenberg model on this
cluster. Therefore, the GS is disentangled, with $C_{ij}=0$,
and one finds that the spin correlations are exactly the same as for
the 1D chain of $N=6$ sites with PBC, i.e., $S_{ij}=-0.4671$.

Orbital fluctuations gradually increase with increasing $\alpha$, but
the above GS remains stable up to $\alpha\simeq 0.10$. For
larger $0.10<\alpha<0.44$ the fluctuations "soften" the orbital state
and allow for local fluctuations along the bonds. As spins are also
involved, this change of the GS is not gradual but occurs as
a quantum transition to the state with degeneracy $d=2$. In this
GS the spin correlations weaken to $S_{ij}\simeq -0.27$ and
the joint spin-and-orbital fluctuations contribute with finite mixed
correlation function, $C_{ij}\simeq -0.12$, see Fig. \ref{fig:hex}(a).

The degeneracy of the GS results in different values of the possible
orbital occupancy at each site --- while the probability of occupying
the orbital $c$ at site 1 is now reduced to $n_{1c}=2/3$ in each
component, the remaining orbital with finite density is either $a$ or
$b$, depending on the actual wave function. Each component
satisfies the local constraint given in Eq. (\ref{const}). As a result,
the average density $n_{1a}=n_{1b}=1/6$ follows from two contributing
eigenfunctions, with either $n_{1a}=1/3$ or $n_{1b}=1/3$, and the third
orbital empty ($n_{1b}=0$ or $n_{1a}=0$). This is marked in Fig.
\ref{fig:hex}(b) by the vertical lines (error bars) which indicate the
range of possible values for the relevant electron density
$n_{1\gamma}$.

In agreement with intuition, when $\alpha=0.5$ and all interorbital
transitions shown in Fig. \ref{fig:hops} have equal amplitude, there
is large orbital mixing which is the most prominent feature in the
GS found in the intermediate regime of $0.44<\alpha<0.63$. It is not
centered at $\alpha=0.5$ as there are less processes which contribute
to direct exchange than to superexchange, the energy gain is lower in
the direct exchange regime and thus this regime which comes next
is narrower. Although both superexchange and direct exchange suggest
AF spin couplings, the orbitals fluctuate here strongly and couple to
the spins. Therefore, AF spin correlation function is again reduced to
$S_{ij}\simeq -0.21$.
Remarkably, both the orbital correlations, $T_{ij}\simeq -0.18$, and
the mixed correlations, $C_{ij}\simeq -0.13$, are also negative, and
the GS is unique ($d=1$). Here all the orbitals contribute equally and
$n_{1\gamma}=1/3$, as seen in the inset of Fig. \ref{fig:hex}(b).
We recognize this state as a prerequisite of the
spin-orbital liquid state which dominates the behavior of the
triangular lattice, as we demonstrate below in Sec. \ref{sec:pbc}.
Actually, there is also some similarity between this fluctuating GS
and the GS of the 1D SU(4) spin-orbital model,
\cite{Ole06} but here the symmetry is lower by construction.

The regime of larger values of $\alpha>0.63$ favors the direct exchange
interactions, supported by pairs of identical orbitals active on an
exchange bond. Having only one orbital flavor active along each bond,
only three bonds may be occupied by spin singlets and contribute with a
direct exchange energy, and there are two distict configurations with
differently distributed orbital occupations along the hexagonal ring,
shown in Fig. \ref{fig:hex}(b). These two distinct GSs (degeneracy
$d=2$) cause again two distributions of the orbital densities, this
time varying between 0 and 1 in the majority of the direct exchange
regime, i.e., for $\alpha>0.7$. After averaging over two degenerate
states, the average occupancy for the orbitals which are active on one
of the bonds originating from each site $i$ is $n_{ia}=n_{ib}=0.5$.
However, there is also a narrow range of $0.63<\alpha<0.7$, where such
fluctuations have a lower amplitude, only between $1/3$ and $2/3$.
This suggests that the orbital fluctuations play still an important
role here and couple weakly since $C_{ij}\simeq 0.05$, in contrast to
$\alpha>0.7$ characterized simply by two distinct orbital configurations
and factorized spin and orbital degrees of freedom, i.e., $C_{ij}=0$.
Indeed, as a result of averaging over two degenerate wave functions in
the GS, see inset of Fig. \ref{fig:hex}(b), one finds at $\alpha=1$
the bond correlations $S_{ij}=-3/8$ and $T_{ij}=1/8$. These
results demonstrate that two orbital configurations are static.

\subsection{Triangular clusters with open boundary conditions}
\label{sec:tri}

In contrast to the hexagonal cluster considered above, triangular
clusters of Fig. \ref{fig:clu}(a) are characterized by frustrated
spin-orbital interactions and contain nonequivalent bonds. Therefore
the case of decoupled spin and orbital dynamics cannot be realized in
the superexchange limit (at $\alpha=0$) in none of the clusters. In
fact, the orbitals try to adjust themselves to the frustrated geometry,
but in general several equivalent configurations contribute and the
obtained results follow from averaging over them. The smallest triangular cluster
N3 was already analyzed in Ref. \onlinecite{Nor08}, we therefore
concentrate here on N6 and N10. Below we discuss first the
bond correlation functions and next explain them by presenting the
electron distribution over $t_{2g}$ orbital states.

\begin{figure}[t!]
\includegraphics[width=8cm]{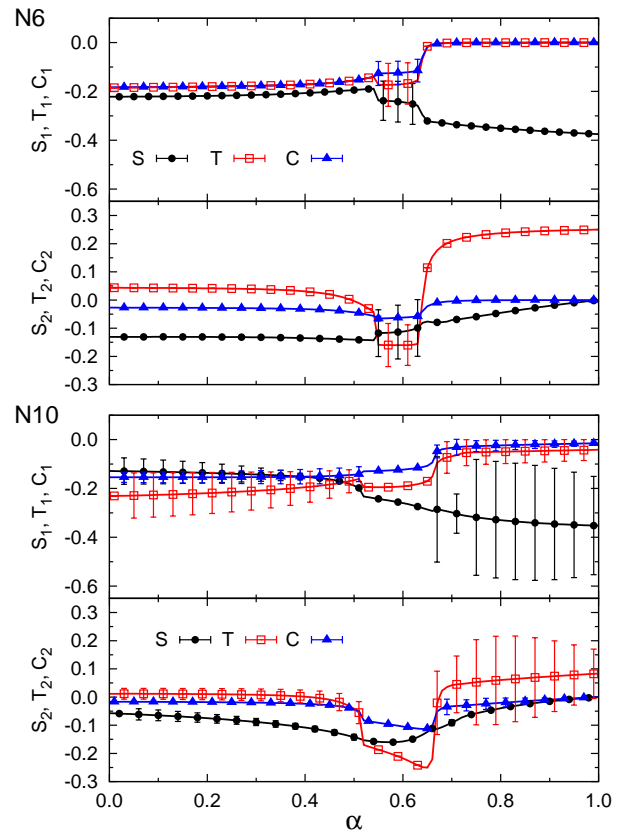}
\caption{(Color online) Bond correlations for triangular clusters for
increasing $\alpha$ in absence of Hund's exchange ($\eta=0$), as obtained
for topologically equivalent bonds $n=1,2$ in triangular clusters
N6 (top) and N10 (bottom) with OBC, see Fig. 2(a): spin
(${\cal S}_n$, circles), orbital (${\cal T}_n$, squares), and
spin--orbital (${\cal C}_n$, triangles) correlations.  }
\label{fig:sst1}
\end{figure}

The intersite spin, orbital, and spin-orbital correlations are
presented in Fig. \ref{fig:sst1} for two classes
of bonds shown in Fig. \ref{fig:clu}(a): a bond involving a corner site
labeled as 1, and an internal bond being close to a corner
labeled as 2. In order to simplify the notation we label the respective
{\it bond} correlation functions for a bond $n=1,2$ by a bond index as
${\cal S}_n$, ${\cal T}_n$, and ${\cal C}_n$, respectively. The data
points indicate three physically different regimes (similar to some
extent to H6 cluster of Sec. \ref{sec:hex}):
(i) the superexchange regime in a range of small values $\alpha\ge 0$,
(ii) an intermediate regime for values of $\alpha$ close to but
typically larger than $\alpha=0.5$, and
(iii) the direct exchange regime far large values of $\alpha$ close to
$\alpha=1$.
The range of values of $\alpha$ for each of these three regimes depends
on the cluster size, but certain common features can easily be
recognized in Fig. \ref{fig:sst1}.
Altogether, the intermediate regime where the electrons are almost
equally distributed and spin-orbital fluctuations are strong becomes
broader by increasing cluster size from N6 to N10.

\begin{table}[b!]
\caption{
Degeneracy of different GSs found in the $d^1$ spin-orbital model Eq.
(\ref{som0}) at the superexchange point
($\alpha=0$), at finite but small $\alpha=\epsilon$, in the
intermediate regime ($\alpha=0.6$), close to ($1-\alpha=\epsilon$) and
at the direct exchange ($\alpha=1$) point for the triangular N3, N6
and N10 clusters shown in Fig. 2(a).
}
\vskip .2cm
\begin{ruledtabular}
\begin{tabular}{cccccc}
cluster  & $\alpha=0$ & $\alpha=\epsilon$ & $\alpha=0.6$
& $\alpha=1-\epsilon$ & $\alpha=1$  \cr
\colrule
  N3  &  6  &  4  &  2  &  8  & 12  \cr
  N6  &  2  &  1  &  2  &  1  &  2  \cr
 N10  &  1  &  2  &  1  &  2  &  6  \cr
\end{tabular}
\end{ruledtabular}
\label{tab:obc}
\end{table}

The clusters N6 and N10 do not have any spin degeneracy and total spin
state is a singlet ${\cal S}=0$ in the entire parameter range. The
degeneracy of N6 cluster (see Table I) follows therefore from the
distribution of three spin singlets over the cluster accompanied by
matching the orbital state in such a way, that the energy of these
singlet states can indeed contribute to the GS energy. The overall
tendency towards singlet spin correlations on the bonds originating
from a corner site 1, see Fig. \ref{fig:clu}(a), is observed in both
N6 and N10 clusters, where ${\cal S}_1< -0.12$ in the entire regime of
$\alpha$. Gradual
localization of spin singlets near corner sites with increasing $\alpha$
is suggested by the decreasing values of ${\cal S}_1$, with the lowest
values reached at $\alpha=1$ in both cases. The superexchange regime
extends up to $\alpha\simeq 0.54$ ($\alpha\simeq 0.51$) for N6 (N10)
cluster, while the direct exchange dominates for $\alpha>0.64$
($\alpha>0.67$), respectively. A small increase of ${\cal S}_1$ in N6
cluster for
$0<\alpha<0.54$ could be understood as a consequence of gradual increase
of orbital fluctuations towards the intermediate regime, where spin and
orbital dynamics start to decouple from each other. This trend is
recognized from simultaneous decrease of spin correlations ${\cal S}_1$
and increase of spin-orbital correlations ${\cal C}_1$. A common
feature for both N6 and N10 clusters in the regime of large $\alpha$
is vanishing spin-orbital correlation function (${\cal C}_1=0$), see
Fig. \ref{fig:sst1}.

\begin{figure}[t!]
\includegraphics[width=8cm]{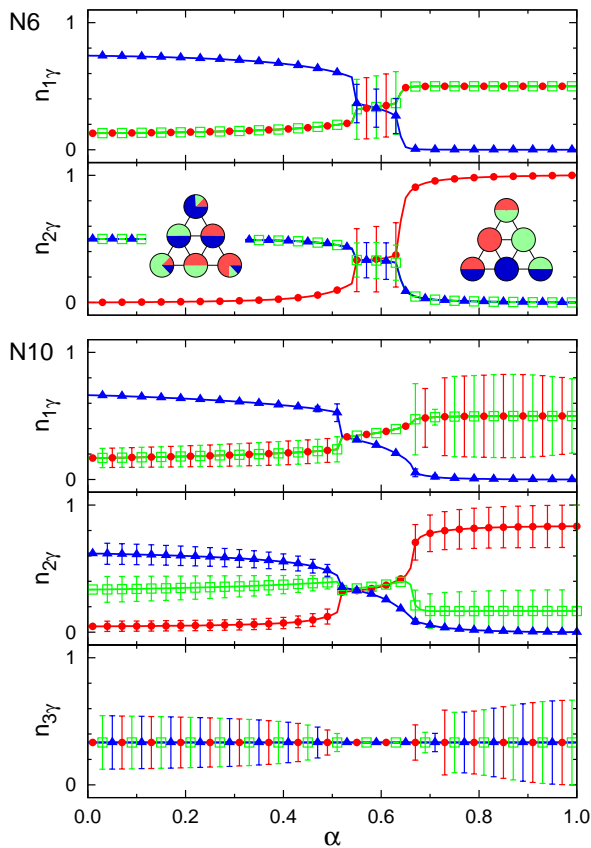}
\caption{(Color online) Orbital occupations $n_{1\gamma}$ in
triangular clusters for increasing $\alpha$ in the absence of Hund's
exchange ($\eta=0$): at sites $i=1,2$ for N6 cluster (top), and sites
$i=1,2,3$ for N10 cluster (bottom), see Fig.~\ref{fig:clu}(a).
Insets show two representative orbital occupancy distributions for the
superexchange (left) and direct exchange (right) regime.
}
\label{fig:nit}
\end{figure}

We remark that also the correlations on bonds labeled as 2 in the
clusters N6 and N10, see Fig. 2(a), have several common features.
First, spin correlations are negative (${\cal S}_2<0$) in the entire
range of $\alpha$ except for $\alpha=1$, where one finds ${\cal S}_2=0$.
We argue that this result manifests frustration
of the internal bonds (labeled 2) in both clusters in contrast to the
bonds originating from a corner (labeled 1). Second, the orbital
correlations ${\cal T}_2$ are positive both for small and large values
of $\alpha$, while in the intermediate regime these correlations are
negative due to large orbital fluctuations. Finally, the spin-orbital
correlations vanish on bonds 2 in both clusters when the direct
exchange point $\alpha=1$ is approached. These results confirm the
observations made above for the bonds 1 originating from a corner that
spin-orbital entanglement is absent at the direct exchange point.

Three different regimes of spin and orbital correlations recognized in
N6 and N10 clusters are characterized by quite different density
distributions of electrons which obey the local constraint Eq.
(\ref{const}). The electron densities, shown in Fig. \ref{fig:nit},
confirm the observations made above by analyzing the intersite
correlations that the GSs of both clusters can be classified as
belonging to three different regimes dominated by:
(i) the superexchange,
(ii) orbital fluctuations, and
(iii) direct exchange.
Different role played by the orbitals $\{a,b,c\}$ along particular
directions in N6 cluster are highlighted by the two insets in
Fig. \ref{fig:nit}. The bonds which originate at a corner site $i=1$
are remarkably similar to each other. The common active orbital $c$ for
the superexchange along both bonds has a large electron density, while
the remaining orbitals may contribute only along one of these bonds. In
the smaller N6 cluster the density distribution at $\alpha=0$ amounts
to $n_{1c}\simeq 0.74$ and $n_{1a}=n_{1b}\simeq 0.13$. This density
distribution cannot be easily deduced by analyzing the degeneracy of
the GS (Table I) and by averaging spin singlet configuration over the
bonds in the N6 cluster which gives instead
$\{n_{1a},n_{1b},n_{1c}\}=\{1/6,1/6,2/3\}$. Therefore, we find here
the first example where the orbital fluctuations play a role and
modify the electron density distribution with respect to the classical
expectations. Nevertheless, the spin singlets are accompanied by
appropriate active orbitals along three isolated bonds, each of them
originating from one corner site of N6 triangle. This is also
confirmed by the density distribution found for the site $i=2$, where
one finds the density of 0.5 for both orbitals active along the bonds
on the triangle edge and 0 for the third orbital which contributes to
the superexchange along the other two triangle edges. To some extent
this is also observed in the larger N10 cluster.

The density distribution at sites $i=1$ and $i=2$ changes towards
more isotropic one for increasing $\alpha$, but the symmetry in the
orbital space is always broken by geometry and the densities differ
somewhat even in the intermediate regime for $\alpha\sim 0.6$ (Fig.
\ref{fig:nit}). For larger values of $\alpha$ one recognizes easily
the situation of singlet spin dimers distributed over the cluster.
Each dimer is based on a single orbital flavor active in the direct
exchange on the particular bond. This state is surprisingly robust
against orbital fluctuations in N6 cluster, where only the orbitals
$a$ and $b$ are occupied at site $i=1$, and $n_{1a}=n_{1b}=0.5$.
Actually, a similar situation is found also in N10 cluster, but
here the GS is degenerate, see Table I, and a range of possible
electron densities $\{n_{1a},n_{1b}\}$ is found instead.
Altogether, we have found that $n_{1c}=0$ in all the considered
triangular clusters when the direct exchange limit is approached.

The distribution of orbital densities at sites labeled $i=2$ in
clusters N6 and N10 in the regime of large $\alpha>0.7$ demonstrates
that indeed these clusters are dominated by the spin singlets touching
a corner site each. It is for this reason that the density $n_{2a}$
approaches $n_{2a}=1$ when $\alpha\to 1$, and the other two orbitals
are empty. In N10 cluster the density $n_{2a}$ is somewhat reduced
due to the geometric constraints and additional frustration introduced
by the total number of five spin singlets which implies several
equivalent states with different orbital density distributions
in the cluster.

The only site in the triangular clusters which recovers full symmetry
in the orbital space is the central $i=3$ site of N10 cluster. Here we
have found that all three orbitals contribute equally in the GS in the
entire range of $\alpha$, with $n_{3\gamma}=1/3$ (Fig. \ref{fig:nit}),
but certain fluctuations around this average value are observed both in
the superexchange ($\alpha<0.52$) and direct exchange ($\alpha>0.66$)
regime. This result may be treated as a precursor of the situation
encountered in the infinite lattice, where the geometrical frustration
favors the disordered state. Such an interpretation is also supported
by the fact that the GS energies per bond (not shown) systematically
increase with increasing cluster size (for fixed $\alpha$) among N3, N6
and N10 clusters, particularly close to superexchange ($\alpha=0$) and
direct exchange ($\alpha=1$) points. We shall investigate this
situation in more detail below
(in Sec. \ref{sec:pbc}) by considering the clusters with PBC.

\subsection{Clusters with periodic boundary conditions}
\label{sec:pbc}

After analyzing the triangular clusters with OBC
in Sec. \ref{sec:tri}, we turn to the clusters N4, N7 and N9 with PBC,
shown in Fig. \ref{fig:clu}(b), which have all the sites equivalent and
are thus representative for the triangular lattice in the thermodynamic
limit. The intersite correlation functions obtained for N4 cluster
were analyzed in Ref. \onlinecite{Nor08}, and we shall concentrate here
on both larger clusters, N7 and N9.

Unlike for the triangular clusters where abrupt transitions between
distinct regimes of particular spin and orbital correlations were found,
one finds here that the intersite spin, orbital and spin-orbital
correlations evolve continuously with increasing $\alpha$ for N7 and N9
clusters, and no distinct regimes with dominating either superexchange
or direct exchange can be identified.
In case of N7 cluster the spin correlations are AF and constant,
$S_{ij}\simeq -0.11$. Note that this value is somewhat higher than the
average obtained by considering randomly distributed spin singlets over
the triangular lattice, i.e., occupying every sixth bond in the lattice
and leading to $\langle S_{ij}\rangle=-0.125$. However, the obtained
smaller value can be justified as follows. In the low-spin phase one
has ${\cal S}=1/2$ total spin, and one can determine the intersite spin
correlations using the following identity:
\begin{equation}
\label{ident}
\vec{\cal S}^2=7\vec{S}_i^2+42\langle\vec{S}_i\cdot\vec{S}_j\rangle\,.
\end{equation}
For this cluster every pair of sites $\{i,j\}$ is a nearest neighbor
pair and forms a bond (due to PBC), so Eq. (\ref{ident}) follows. The
value $\langle\vec{S}_i\cdot\vec{S}_j\rangle=-3/28=-0.107$ obtained
from it is much reduced from the classical AF spin correlations $-1/4$
in the N\'eel state on a square lattice, which is a consequence of high
frustration of the triangular lattice. Note that spin correlations are
here substantially reduced by the geometrical frustration, and seem not
to be further hindered by orbital fluctuations, unlike in the 1D SU(4)
model,\cite{Fri98} where the coupling to the orbital correlations is
crucial and reduces spin correlations although the geometrical
frustration is absent.

\begin{figure}[t!]
\includegraphics[width=8cm]{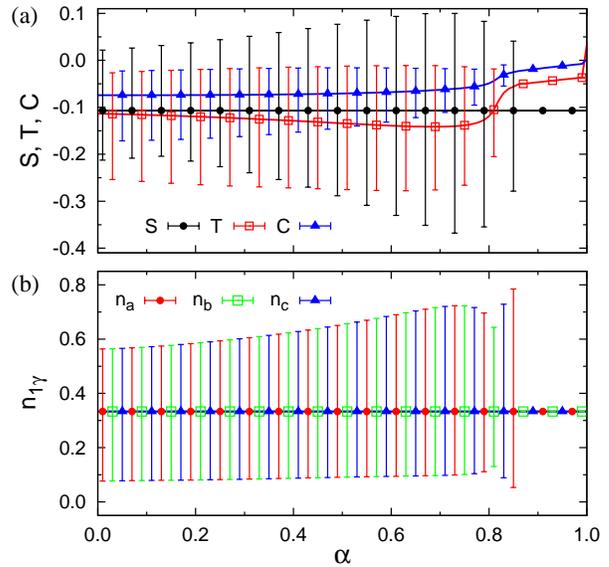}
\caption{(Color online)
Evolution of the frustrated GS for N7 cluster with PBC shown in
Fig. \ref{fig:clu}(b) with increasing $\alpha$ in the absence of Hund's
exchange:
(a) intersite bond correlations --- spin ($S_{ij}$, circles), orbital
($T_{ij}$, squares), and spin-orbital ($C_{ij}$, triangles);
(b)~orbital occupations $n_{1\gamma}$ per site (at a representative
site $i=1$).  }
\label{fig:n7}
\end{figure}

For the values of $\alpha<0.8$ the orbital correlations $T_{ij}$ are
even a bit lower than the spin ones, and decrease somewhat with
increasing $\alpha$ for $\alpha<0.6$. We emphasize that both spin and
orbital correlations are here negative, so the present spin-orbital
model on the triangular lattice violates the Goodenough-Kanamori rules
that these correlations should be complementary.\cite{Goode} This is
also reflected in the finite spin-orbital correlations
$C_{ij}\simeq -0.07$ which indicate that entangled states play an
important role in this parameter range, see Fig. \ref{fig:n7}. When
$\alpha$ increases further, however, the orbital state is reorganized
and the orbital correlations rapidly decrease. At the same time the
mixed spin-orbital correlations also decrease suggesting gradual
disentanglement of spin and orbital degrees of freedom. Finally, at
$\alpha=1$ the spins decouple from the orbitals, which agrees with the
GS consisting of uncoupled spin singlets distributed over the
triangular lattice.\cite{Jac07}

As all the orbitals are equivalent, the average occupancy is
$n_{i\gamma}=1/3$, see Fig. \ref{fig:n7}. However, apart from the spin
degeneracy due to the total spin state ${\cal S}=1/2$ realized for the
cluster with odd $N=7$ sites, one finds that several equivalent orbital
configurations contribute, with one of the orbitals occupied by three
electrons and the other two by two each. For this reason we have found
degeneracy $d=6\times 2=12$, see Table II and large fluctuations of the
density distribution, typically between 0.1 and 0.6 for low values of
$\alpha<0.5$, and increasing towards $\alpha=0.85$. Next the quantum
nature of the GS makes it only spin degenerate and $d=2$ for
$\alpha>0.85$ and no fluctuations in the orbital occupancies were found.
Finally, in the direct exchange limit $\alpha=1$ independent dimer
distributions over the cluster dominate and determine large orbital
degeneracy $d=147\times 2$ of the GS. This degeneracy due to the orbital
distribution over the cluster can be understood
as given by 3 possibilities of having one dominating orbital flavor,
7 positions of this extra flavor in the cluster, and 7 possible
distributions of singlets over the cluster, i.e.,
$3\times 7\times 7=147$.

\begin{table}[b!]
\caption{
Degeneracy $d$ of the GSs found for the $d^1$ spin-orbital model Eq.
(\ref{som0}) at the superexchange point ($\alpha=0$), in the
intermediate regime ($0<\alpha<1$) and at the direct exchange point
($\alpha=1$) obtained for N7 and N9 clusters with PBC. The first factor
in $d$ gives the orbital degeneracy which is multiplied by spin
degeneracy 2 for the GS with ${\cal S}=1/2$ total spin.
}
\vskip .2cm
\begin{ruledtabular}
\begin{tabular}{ccccc}
& cluster  &  N7  &  N9  &   \cr
\colrule
& $0\le\alpha<0.26$  &  $6\times 2$  &  $6\times 2$  &   \cr
& $0.26<\alpha<0.41$ &  $6\times 2$  &  $4\times 2$  &   \cr
& $0.41<\alpha<0.85$ &  $6\times 2$  &  $2\times 2$  &   \cr
& $0.85<\alpha<1$    &  $1\times 2$  &  $6\times 2$  &   \cr
&   $\alpha=1$       & $147\times 2$ & $756\times 2$ &   \cr
\end{tabular}
\end{ruledtabular}
\label{tab:pbc}
\end{table}

\begin{figure}[t!]
\includegraphics[width=8cm]{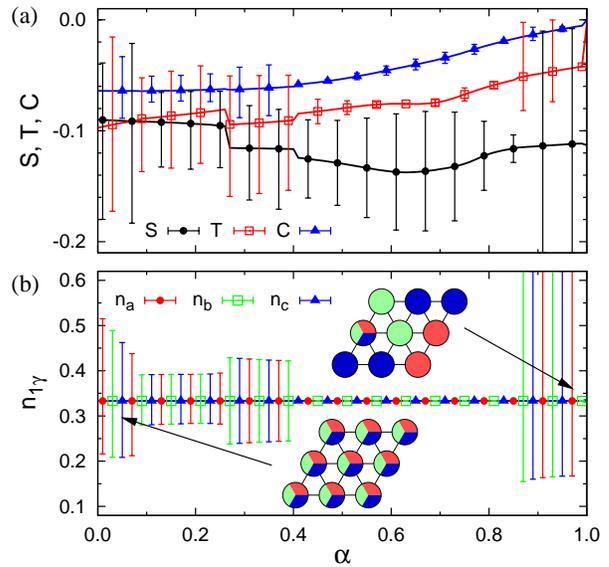}
\caption{(Color online)
Evolution of the frustrated GS for N9 cluster with PBC
shown in Fig. \ref{fig:clu}(b) with increasing $\alpha$ in the absence
of Hund's exchange:
(a)~intersite bond correlations --- spin ($S_{ij}$, circles), orbital
($T_{ij}$, squares), and spin-orbital ($C_{ij}$, triangles);
(b)~orbital occupations $n_{1\gamma}$ per site (at a representative
site $i=1$).
Insets show typical orbital patterns in the superexchange
($\alpha=0$) limit and direct exchange ($\alpha=1$) limit. }
\label{fig:n9}
\end{figure}

The second cluster with PBC contains $N=9$ sites and is better designed
to study the present model as the degeneracy of the GS is expected to
be lower (except at $\alpha=1$). In this case electrons can be equally
distributed over the orbitals, and three electrons occupy each of them.
Fluctuations over different values of bond orbital correlations $T_{ij}$
and differently occupied orbital states, see Figs. \ref{fig:n9}(a) and
\ref{fig:n9}(b), cannot be avoided within the cluster, but they
are typically smaller than those found for N7 cluster. All average
values for the intersite correlations are rather similar to those found
in N7 cluster for small $\alpha$, with $S_{ij}\simeq -0.90$,
$T_{ij}\simeq -0.97$, and $C_{ij}\simeq -0.64$ at $\alpha=0$.
Here all the orbital states are equally populated as shown by the inset
in Fig. \ref{fig:n9}, but undergo local fluctuations, seen both in the
intersite correlation functions $\{S_{ij},T_{ij},C_{ij}\}$ and in the
electron densities $\{n_{ia},n_{ib},n_{ic}\}$.

The range of fluctuations in bond correlations is reduced but stays
finite up to $\alpha\simeq 0.41$. Here the orbital degeneracy of the GS
is first 6 up to $\alpha=0.26$, and next drops to 4, see Table II. This
regime is followed by a qualitatively new situation in the intermediate
regime of $\alpha$ values (compared to N7 cluster), with no
fluctuations in the orbital distribution when $0.41<\alpha<0.85$. This
regime is characterized by low degeneracy 2 in the orbital space. In
this case the orbitals undergo strong local quantum fluctuations but
their distribution in the cluster does not change, as seen in the
stable density distribution, with $n_{1\gamma}=1/3$. This regime can be
identified as dominated by orbital fluctuations due to the mixing terms
${\cal H}_m$ Eq. (\ref{Hm}), with gradual suppression of spin-orbital
fluctuations under increasing $\alpha$ seen in the reduced values of
$|C_{ij}|$.

Finally, when the orbital mixing terms sufficiently decrease, one finds
that both spin and orbital correlations fluctuate stronger above
$\alpha=0.85$ which follows from a random electron distribution over the
available orbital states. In contrast to the superexchange regime with
equally distributed electrons over the orbital flavors, the
representative state shown in the inset in Fig. \ref{fig:n9}, is
dominated by $c$ orbitals.
Equivalent configurations can be obtained by cyclic permutations of the
density distribution in the $\{a,b,c\}$ orbitals, and changing favored
dimer direction in the cluster. For this reason degeneracy of the
GS increases by a factor of 3, reflecting three possible with
states dominated by one orbital flavor, see Table II. The direct
exchange limit is again special, and characterized by large orbital
degeneracy 756.

\begin{figure}[t!]
\includegraphics[width=7.5cm]{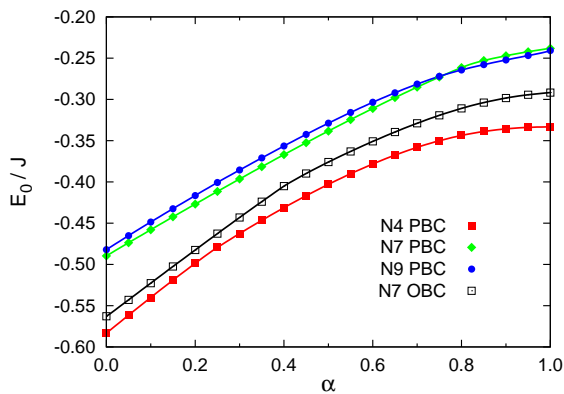}
\caption{(Color online) Ground state energy $E_0$ per bond for
clusters N4 (filled squares), N7 (diamonds), and N9 (circles) with PBC
conditions for increasing $\alpha$. For a comparison the result
obtained for N7 cluster with OBC is shown by empty squares.
Parameter: $\eta=0$. }
\label{fig:epbc}
\end{figure}

The GS energies obtained for N4, N7 and N9 clusters with the PBC
(Fig. \ref{fig:epbc}) exhibit a monotonous increase with increasing
$\alpha$. The energy is the lowest for the smallest N4 cluster, and
increases when the cluster size is increased to N7 or N9, indicating
frustrated interactions. However, the energies obtained for N7 or N9
clusters are very close to each other which we take as an evidence that
these clusters are already representative for the situation in the
thermodynamic limit. We remark that the energy obtained for N7 cluster
with OBC is lower than that found for the same cluster size with PBC
which is again consistent with strongly frustrated spin-orbital
interactions on the triangular lattice. This resembles the situation
in spin systems with frustrated interactions, where one expects that
clusters with OBC would give lower energy per bond than the ones with
PBC as then additional breaking of symmetry is possible in the GS.

\section{The model at large Hund's exchange}
\label{sec:his}

\subsection{Orbital model for the FM phase}
\label{sec:fm}

Until now we considered the spin-orbital model Eq. (\ref{som}) in the
regime of no Hund's exchange, $\eta=0$. Finite Hund's exchange is
responsible for the competition of FM with AF spin interactions in the
perovskite Mott insulators described by similar spin-orbital models, and
the orbital order is usually modified in a particular way. Well known
examples are the $A$-AF phase in LaMnO$_3$,\cite{Fei99} or $C$-AF phase
in LaVO$_3$.\cite{Kha01} The present and earlier\cite{Nor08} study of
the spin-orbital $d^1$ model on the triangular lattice, however, suggest
that orbital disorder is favored in this geometrically frustrated
lattice. Having no orbital order excludes {\it de facto} an intermediate
phase (at intermediate values of $\eta$) with coexisting FM and
AF interactions in the present case. This
is confirmed by the exact diagonalization of finite clusters, at least
in the considered case when all the interactions along three bond
directions are equivalent. Thus, there are two problems to be addressed
in the theory:
(i) the nature of orbital correlations in the FM phase in the range of
large values of $\eta$, and
(ii) the phase diagram in the $(\alpha,\eta)$ plane.
The first of these questions is easier to answer and we consider it
below, the second one will be discussed in Secs. \ref{sec:phd} and
\ref{sec:enta}.

The exchange interactions in ${\cal H}$ given by Eq. (\ref{som})
simplify in the FM phase, when only excitations to high-spin states
contribute and all the low-spin terms vanish, i.e.,
$\langle\vec{S}_i\cdot\vec{S}_j-\frac14\rangle\equiv 0$.
The Hamiltonian reduces then to the orbital model,
\begin{eqnarray}
\label{Hfm}
{\cal H}_{\rm orb} \!\!\! & = & \! -\frac{1}{4}Jr_1\!
\sum_{\langle ij \rangle \parallel \gamma}\! \Big\{ -(1-\alpha)\Big[
2A_{ij}^{(\gamma)} + (n_{i\gamma} + n_{j\gamma}) - 2 \Big]
\nonumber \\
&+&\!\sqrt{\alpha(1-\alpha)}\,\Big(
           T_{i\mu}^+T_{j\nu}^+ +T_{i\nu}^+ T_{j\mu}^+ +
           T_{i\nu}^-T_{j\mu}^- +T_{i\mu}^- T_{j\nu}^- \Big)
\nonumber \\
&+&\! \alpha\, \Big[ n_{i\gamma} (1 - n_{j\gamma})
                 +  (1 - n_{i\gamma}) n_{j\gamma}\Big]\Big\}\,.
\end{eqnarray}
In the superexchange regime ($\alpha\simeq 0$) it favors pairs of
different orbitals, both not oriented along the considered bond, and
in the direct exchange regime ($\alpha\simeq 1$) --- pairs of
different orbitals, one oriented along the bond and the other not.

Complete information about the orbital correlations in the parameter
regime where the FM phase is stable may be obtained by considering the
following orbital projection operators for a bond
$\langle ij \rangle\parallel\gamma$:
\begin{eqnarray}
\label{orbipro}
 P_{ij}^{(\gamma)} & = & \langle n_{i\gamma} n_{j\gamma} \rangle\,, \\
 Q_{ij}^{(\gamma)} & = & \langle n_{i\gamma} (1 - n_{j\gamma}) \rangle
     +   \langle (1 - n_{i\gamma}) n_{j\gamma} \rangle\,, \\
 R_{ij}^{(\gamma)} & = & \langle (1 - n_{i\gamma}) (1 - n_{j\gamma}) \rangle\,.
\end{eqnarray}
Here the operator $n_{i\gamma}$ stands for the electron density in the
orbital oriented along the bond $\langle ij \rangle$ at site $i$, while
$(1-n_{i\gamma})$ is the complementary electron density in the two
remaining orbitals, as given by the local constraint Eq. (\ref{const}).
The above projection operators may be treated as probabilities to
encounter a given orbital configuration, and they obey the usual condition,
\begin{equation}
\label{pqr}
P_{ij}^{(\gamma)} + Q_{ij}^{(\gamma)} + R_{ij}^{(\gamma)} = 1\,.
\end{equation}
Knowing these probabilities allows us for a complete characterization
of the orbital state on a representative bond
$\langle ij\rangle\parallel c$.

\begin{figure}[t!]
\includegraphics[width=8cm]{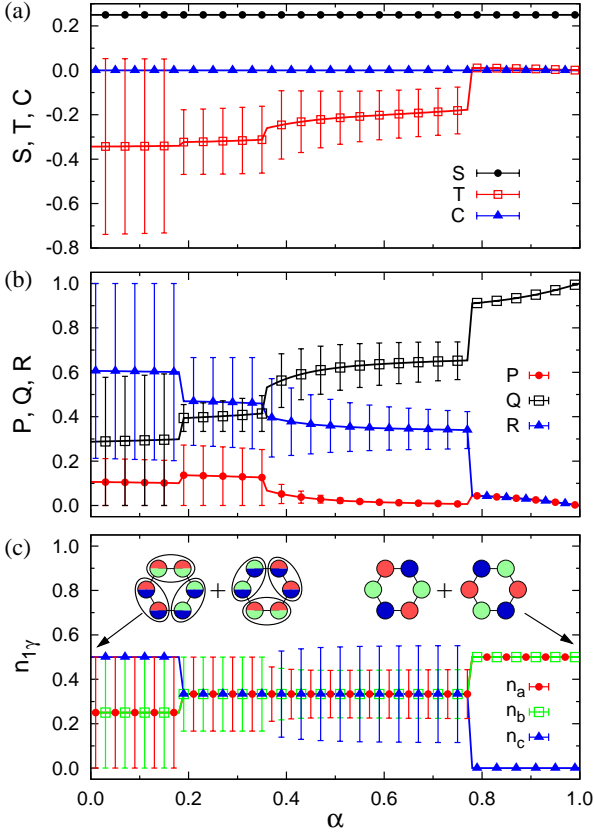}
\caption{(Color online)
Evolution of the orbital state in the FM GS of the hexagonal H6 cluster
with OBC, as found at large $\eta=0.2$ for increasing $\alpha$:
(a) spin $S\equiv S_{ij}$, orbital $T\equiv T_{ij}$ and
spin-orbital $C\equiv C_{ij}$ bond correlations;
(b)~orbital bond projection operators $P\equiv P_{ij}^{(\gamma)}$
(circles), $Q\equiv Q_{ij}^{(\gamma)}$ (squares),  and
$R\equiv R_{ij}^{(\gamma)}$ (triangles);
(c)~orbital electron densities $n_{1\gamma}$
on the most left cluster site $i=1$. Insets in (c) show two
equivalent states in the superexchange ($\alpha=0$) and direct exchange
($\alpha=1$) limit.}
\label{fig:g6fm}
\end{figure}

First we present distinct differences between the distribution of
occupied orbitals between the superexchange ($\alpha\simeq 0$) and
direct exchange ($\alpha\simeq 1$) regime found for the FM hexagonal
H6 cluster. Here the orbital bond correlations change from
$T_{ij}\simeq -0.34$ at $\alpha=0$ to $T_{ij}\simeq 0$ at $\alpha=1$,
while the spins are FM ($S_{ij}=0.25$) and disentangled from the
orbital state ($C_{ij}=0$), see Fig. \ref{fig:g6fm}(a). This latter
feature is common for FM states in other clusters as well (see below)
and the Goodenough-Kanamori rule\cite{Goode} is obeyed, i.e., FM
spin correlations are accompanied by negative orbital correlations
which indicate that orbitals show a tendency towards alternating
orbital order.

The FM state
in the superexchange regime ($\alpha\simeq 0$) is stabilized by pairs
of orbitals which do not include the one oriented along the considered
bond. This is indicated by the inset in Fig. \ref{fig:g6fm}(c), where
two possible configurations with pairs of active orbitals on the bonds
of H6 cluster, forming almost orbital singlets, are shown.
In this case the largest
average projection operator is $\langle R_{ij}^{(\gamma)}\rangle$, see
Fig. \ref{fig:g6fm}(b), and the orbital density is the largest for the
$c$ orbital which participates in the singlets for both possible
cluster coverings, i.e., $n_{1c}=0.5$ and $n_{1a}=n_{1b}=0.25$.

The intermediate regime extends in the FM H6 cluster from $\alpha=0.19$
to $\alpha=0.77$, where the orbital densities are the same in each
orbital, $n_{1\gamma}=1/3$, see Fig. \ref{fig:g6fm}(c). However, the
orbital correlations on the bonds are not constant, but gradually
change towards pairs of one $\gamma$ orbital accompanied by a different
orbital on each bond $\langle ij\rangle\parallel\gamma$ with
increasing $\alpha$, as depicted by the projection operator
$\langle Q_{ij}^{(\gamma)}\rangle$ in Fig. \ref{fig:g6fm}(b).
Simultaneously the orbital correlation $T_{ij}$ is gradually reduced.
Finally, for $\alpha>0.77$ the orbital distribution changes to pairs
of orbitals, one oriented along the bond and the other not, favored
in the direct exchange regime. Once again, there are two possible
distributions of orbitals over the H6 cluster, and this gives for
site $i=1$ two contributing densities: $n_{1a}=n_{1b}=0.5$, while
the orbital which participates in the direct exchange only on two
horizontal bonds is empty, i.e., $n_{1c}=0$,
see Fig. \ref{fig:g6fm}(c).

\begin{figure}[t!]
\includegraphics[width=8cm]{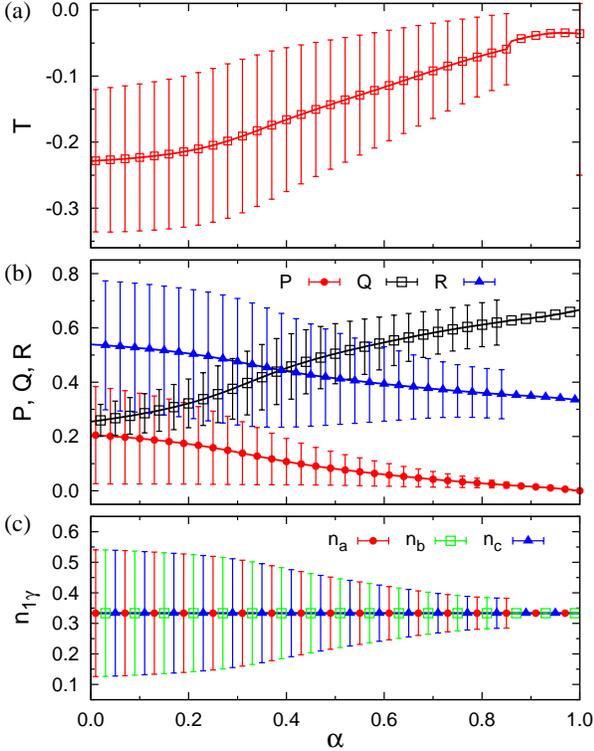}
\caption{(Color online)
Evolution of the orbital state in the FM GS of N7 cluster for
increasing $\alpha$, found with PBC at large $\eta=0.2$:
(a)~bond orbital correlations $T\equiv T_{ij}$;
(b)~orbital occupation correlations $P\equiv P_{ij}^{(\gamma)}$
(circles), $Q\equiv Q_{ij}^{(\gamma)}$ (squares),  and
$R\equiv R_{ij}^{(\gamma)}$ (triangles);
(c)~orbital electron densities $n_{1\gamma}$. }
\label{fig:ferro}
\end{figure}

The above transparent picture of the occupied orbitals in different
$\alpha$ regimes for H6 cluster is strongly modified in the GS of
the symmetric N7 cluster by frustration of spin-orbital interactions
and by the absence of symmetry breaking in the orbital space due to the
PBC, see Fig. \ref{fig:ferro}. At $\alpha=0$ one finds negative orbital
correlation function $T_{ij}\simeq -0.23$, see Fig. \ref{fig:ferro}(a).
The orbital correlations are quite well developed
here as they are not hindered by spin fluctuations, as it was the case
at $\eta=0$ (Fig. \ref{fig:n7}). With increasing $\alpha$ the
orbital correlations are gradually reduced, and reach the limiting
value $T_{ij}=-0.0357$ at $\alpha=1$ which corresponds to randomly
distributed dimers consisting of a pair of active and inactive orbital
in the direct exchange over the bonds of N7 cluster.

Two distinct regimes: one with fluctuating values of $T_{ij}$ for
$\alpha<0.85$, and the other one with uniquely determined $T_{ij}$ for
$\alpha>0.85$ correspond to the degenerate and nondegenerate GS of the
N7 cluster, respectively.
Similar fluctuating results for the average projection operators
$\{P_{ij}^{(\gamma)},Q_{ij}^{(\gamma)},R_{ij}^{(\gamma)}\}$ and for the
orbital densities $\{n_{1a},n_{1b},n_{1c}\}$ were found in the range of
$\alpha<0.85$, see Figs. \ref{fig:ferro}(b) and \ref{fig:ferro}(c).
In the regime of small $\alpha<0.4$ configurations with two orbitals
active in the superexchange processes are the most probable ones, i.e.,
$R_{ij}^{(\gamma)}>P_{ij}^{(\gamma)}$ and
$R_{ij}^{(\gamma)}>Q_{ij}^{(\gamma)}$. On the contrary, when
$\alpha>0.4$, pairs of orbitals, one active and the other one inactive
in the direct exchange processes, dominate, i.e.,
$Q_{ij}^{(\gamma)}>R_{ij}^{(\gamma)}$ and $P_{ij}^{(\gamma)}\to 0$
when $\alpha\to 1$. Here the results suggest a single quantum state
with equally distributed orbital flavors over the cluster,
$n_{1\gamma}=1$.

We have verified that the above evolution of the orbital state in N7
cluster is representative for the present triangular lattice in the FM
regime by considering two larger clusters with PBC (not shown):
(i) N12 cluster obtained by adding 5 sites to N7, e.g. on the right
hand side and on top, and (ii) a star-like N13 cluster obtained by
adding a triangle to each side of N7 cluster. Both clusters have
all sites equivalent and may be used to cover the lattice (in the
second case two equivalent coverings differ by chirality). In both
cases we found that the orbital bond correlations $T_{ij}$ increase
from $T_{ij}\simeq -0.23$ at $\alpha=0$ to $T_{ij}\simeq 0$ at
$\alpha=1$. The occupied orbitals give again (as for N7) large
$R_{ij}^{(\gamma)}\simeq 0.5$ at the superexchange limit $\alpha=0$.
It decreases with increasing $\alpha$, and one finds instead large
$Q_{ij}^{(\gamma)}\simeq 2/3$ at the direct exchange case $\alpha=1$.
Interestingly, the orbital fluctuations shown by the vertical lines are
reduced in both cases, and they do not occur at all in N12 cluster
both for $\alpha<0.62$ and for $\alpha>0.81$.

\subsection{Phase diagrams}
\label{sec:phd}

Perhaps the most intriguing question concerning the GS of the
spin-orbital model Eq. (\ref{som}) is the phase diagram and the way
the FM state occurs as a function of both model parameters, $\alpha$
and $\eta$. Having no possibility to access the phase diagram in the
thermodynamic limit (see also Sec. \ref{sec:enta}), we shall
concentrate here on a few representative clusters trying to extract
the common and generic features of the low-to-high spin transition.

\begin{figure}[t!]
\includegraphics[width=8cm]{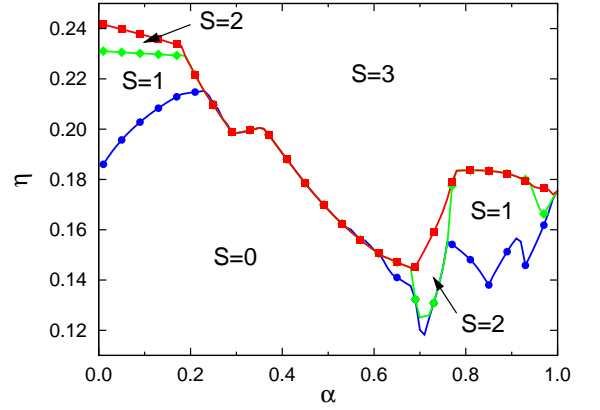}
\caption{(Color online) Phase diagram in the $(\alpha,\eta)$ plane
as obtained for the hexagonal H6 cluster with OBC.
The total spin of the GS is indicated in the
transition regime from the singlet phase (${\cal S}=0$)
to the high-spin (${\cal S}=3$) phase.}
\label{fig:phdhex}
\end{figure}

The hexagonal H6 cluster, with low-spin and high-spin states shown in
Figs. \ref{fig:hex} and \ref{fig:g6fm}, serves here as an example of
unfrustrated geometry. The transition to the high-spin
${\cal S}=3$ state is then gradual and passes through intermediate
${\cal S}=1$ and (in some cases) ${\cal S}=2$ states when $\alpha$ is
close to 0 or 1, see Fig. \ref{fig:phdhex}. In contrast, for the
intermediate values of $0.22<\alpha<0.62$ the transition takes place
directly between the ${\cal S}=0$ and ${\cal S}=3$ states.
We recall that in the regime of intermediate $\alpha$ values orbital
fluctuations play an important role, and they also couple to spin
fluctuations. Therefore, this behavior in the phase diagram reflects
the stabilizing role of the joint spin-orbital fluctuations in the
singlet phase, ${\cal S}=0$. When spin value increases, such
fluctuations are partly damped and therefore intermediate spin values
are here not realized.

An interesting result was obtained for H6 cluster in the
superexchange regime (at $\alpha\simeq 0$). Here the transition from
the singlet to the ${\cal S}=1$ state occurs at $\eta=0.186$. It is
followed by the transition to the ${\cal S}=2$ state at $\eta=0.231$,
and the final transition to the spin-polarized ${\cal S}=3$ state
takes place only at $\eta=0.242$. Here the phases with intermediate
spin values arise as a consequence of spin fluctuations that stabilize
them in between the low-spin and high-spin states. A similar situation
was found close to but not at the direct exchange point $\alpha=1$.
In fact, spin fluctuations in the direct exchange limit concern only
pairs of spins in the singlet ${\cal S}=0$ phase, so a singlet-triplet
transition for a single bond induces here the global transition to the
${\cal S}=3$ phase. Altogether, the transition to the high-spin phase
occurs here at $\eta_c\simeq 0.188$, in agreement with the classical
expectation, see also Eq. (\ref{etac}) in Sec. \ref{sec:enta}, and
this critical value is enhanced in the superexchange dominated regime
due to spin-orbital fluctuations which are here stronger in the singlet
phase.

\begin{figure}[t!]
\includegraphics[width=8cm]{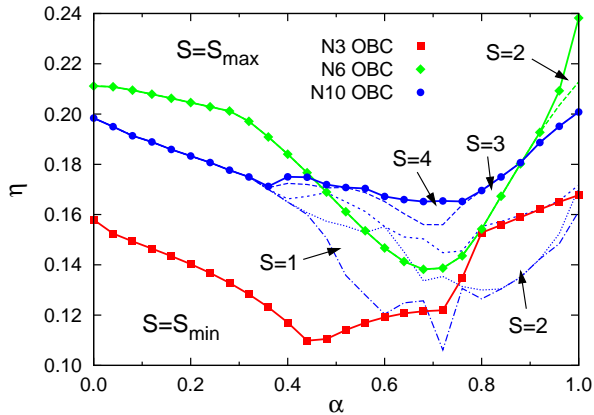}
\caption{(Color online) Phase diagram in the $(\alpha,\eta)$ plane
as obtained for triangular clusters with OBC: N3
(squares),\cite{notec1} N6 (diamonds) and N10 (circles).
The phase diagrams of N6 and N10 clusters contain also intermediate
spin phases and the corresponding phase boundaries are shown as
dashed lines (for the N6 cluster only near $\alpha=1$).}
\label{fig:phdtri}
\end{figure}

The phase diagrams obtained for three triangular clusters, N3, N6, and
N10, are shown in Fig. \ref{fig:phdtri}. As for the hexagonal H6
cluster, the transition to the high-spin phase occurs for
$\eta\simeq 0.18$, but the critical value of $\eta$ is larger for two
bigger clusters (N6 and N10) than for the smallest N3 triangular
cluster.\cite{notec1}
One finds here once again a signature of the stabilizing role played by
orbital fluctuations in the low-spin phase. Particularly in the regime
dominated by the superexchange, the onset of the high-spin phase occurs
at values of $\eta$ which are higher than at $\alpha\simeq 0.5$ and
increase from N3 ($\eta=0.158$) to larger clusters ($\eta=0.211$ and
$\eta=0.198$ for N6 and N10 clusters). In all three cases the transition
occurs to the phases with maximal spin and no intermediate phases in
between.

Phases with intermediate spin values occur for two larger triangular
clusters in the intermediate and direct exchange regime. We suggest
that they are stabilized by orbital fluctuations which can couple to
only partly polarized spin subsystem and provide certain energy gain.
While this feature is general, the actual range of stability of the
phases with intermediate spin values depends on the cluster size. At
$\alpha=1$ one finds for N6 a transition to ${\cal S}=3$ phase at a
rather large value of $\eta=0.238$, which demonstrates here
particular stability of states with lower spin values, where more
energy can be gained due to quantum fluctuations for certain orbital
arrangements. However, in a larger N10 cluster this transition occurs
at a lower value of $\eta=0.202$.

\begin{figure}[t!]
\includegraphics[width=8cm]{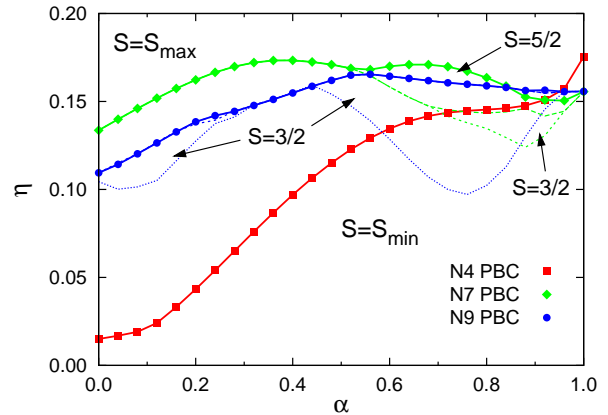}
\caption{(Color online)
Phase diagram in the $(\alpha,\eta)$ plane as obtained for
clusters with PBC:
N4 (squares),\cite{notec2} N7 (diamonds) and N9 (circles).
Ranges of stability of intermediate spin phases for N7 and
N9 clusters are shown by thin lines and arrows. }
\label{fig:phdpbc}
\end{figure}

When the phase diagrams obtained for the triangular clusters (Fig.
\ref{fig:phdtri}) are compared with those for the clusters N7 and N9
with PBC (Fig. \ref{fig:phdpbc}), one finds that the transition to the
high-spin states occurs in general for somewhat lower values of $\eta$
in the latter case. An extreme case here is N4 cluster with PBC,
where a rather small value of $\alpha\simeq 0.02$ suffices to
destabilize the singlet phase at $\alpha=0$. This peculiar result
follows from the small size of this cluster which allows one to
accommodate only two spin singlets when the orbital state is constrained
to the pairs of orbitals supporting the superexchange processes, while
in the high-spin ${\cal S}=2$ state these constraints are released and
the orbital fluctuations of different kind stabilize it. In contrast,
a similar critical value in the range of $0.15<\eta<0.175$ is found for
the low-to-high spin transition in all the clusters N4, N7, and N9 with
PBC at $\alpha=1$. This common feature which is almost independent of
the cluster size suggests that static dimer configurations dominate in
this case not only for the low-spin phase,\cite{Jac07} but also for the
high-spin phase, where pairs of different orbitals are stable on
individual bonds, see Fig. \ref{fig:ferro}.

The proximity of critical values of $\eta$ found for N7 and N9 clusters
is very encouraging and suggests that one may expect the FM phase for
$\eta>0.16$ in the thermodynamic limit, independently of the ratio of
superexchange to direct exchange (i.e., on the actual value of
$\alpha$). In both
clusters we have also found a range of stability for the phases with
intermediate value of total spin which suggests that this transition is
likely to be continuous, via weakly polarized FM states, also in the
thermodynamic limit. It is remarkable, however, that intermediate spin
states do not occur for the direct exchange interactions at $\alpha=1$
which demonstrates once again that quantum fluctuations do not play an
important role in this case, and simple configurational averaging
over available orbital dimer configurations on the lattice suffices to
understand the magnetic transition described here.

\section{Spin-orbital entanglement}
\label{sec:enta}

\subsection{Disentangled spin-orbital interactions}
\label{sec:dis}

When analyzing the phase diagrams for the clusters with PBC, we have
emphasized the role played by orbital and spin-orbital fluctuations,
as well as spin-orbital entanglement in the GS. Now we
shall present additional data to support this claim. In order to
address this question we introduce an approximate treatment of the
spin-orbital Hamiltonian which uses MF decoupling of spin
and orbital variables on the bonds.\cite{Sir08} Focusing on the
magnetic interactions which concern here $S=1/2$ quantum spins coupled
by the SU(2) symmetric interaction, we rewrite the $d^1$ spin-orbital
model Eq. (\ref{som}) in a general form,\cite{Ole05}
\begin{equation}
\label{somjk}
{\cal H} = \sum_{\langle ij \rangle \parallel\gamma} \left\{
{\hat {\cal J}}_{ij}^{(\gamma)} \left( {\vec S}_i \cdot {\vec S}_j
\right) + {\hat {\cal K}}_{ij}^{(\gamma)} \right\}\,,
\end{equation}
where the operators ${\hat {\cal J}}_{ij}^{(\gamma)}$ and
${\hat {\cal K}}_{ij}^{(\gamma)}$ contain orbital pseudospin operators
for a bond $\langle ij\rangle$ along the direction $\gamma$.
This form is helpful as one can deduce the values of spin exchange
constants directly from it when the orbital operators are replaced by
their averages in a particular (ordered or disordered) state. In
particular, it helped to understand the origin of magnetic interactions
in LaMnO$_3$,\cite{Fei99} where such a decoupling scheme may be well
justified and gives predictions for the optical spectral weights
\cite{Ole05} that agree with the experimental data.\cite{Kov10}

Here we introduce the MF procedure to simplify Eq. (\ref{somjk}) as
follows,
\begin{eqnarray}
\label{sommf}
{\cal H}_{\rm MF} &=& \sum_{\langle ij \rangle \parallel\gamma}
\left\langle
{\hat {\cal J}}_{ij}^{(\gamma)}\right\rangle
\left( {\vec S}_i \cdot {\vec S}_j\right) \nonumber \\
&+& \sum_{\langle ij \rangle \parallel\gamma} \left\{
{\hat {\cal J}}_{ij}^{(\gamma)} \left\langle {\vec S}_i\cdot {\vec S}_j
\right\rangle + {\hat {\cal K}}_{ij}^{(\gamma)} \right\} \nonumber \\
&-& \sum_{\langle ij \rangle \parallel\gamma}
\left\langle {\hat {\cal J}}_{ij}^{(\gamma)}\right\rangle
\left\langle {\vec S}_i\cdot {\vec S}_j\right\rangle
\,.
\end{eqnarray}
The first term in Eq. (\ref{sommf}) is the spin model as introduced in
Ref. \onlinecite{Ole05}, the second one is a purely orbital model,
while the last one is a double counting correction term for the
spin-orbital part of the Hamiltonian Eq. (\ref{somjk}). As an example
we consider the orbital operator ${\hat {\cal J}}_{ij}^{(c)}$  which
stands as a coefficient of the Heisenberg spin interaction for a bond
$\langle ij\rangle$ along the $c$ axis,
\begin{eqnarray}
\label{jorbi}
{\hat {\cal J}}_{ij}^{(c)}\!\! &=&\!
J(1-\alpha)\left\{\frac{r_1-r_2}{2}
\left[\frac12\left(n_{ic}+n_{ic}\right)-1\right]\right. \nonumber\\
&+&\!\!\left.\frac{r_1+r_2}{2}\left(T_{ic}^+T_{jc}^+ + T_{ic}^-T_{jc}^-
     -2T_{ic}^z T_{jc}^z + \!\frac12 n_i^{(c)}n_j^{(c)}\!\right)\right.
\nonumber\\
&-&\!\!\left.\frac{r_2-r_3}{2}\left(T_{ic}^+T_{jc}^- + T_{ic}^-T_{jc}^+
     -2T_{ic}^z T_{jc}^z + \!\frac12 n_i^{(c)}n_j^{(c)}\!\right)\!\right\}
\nonumber\\
&+&\!J\sqrt{\alpha(1-\alpha)}\;\frac{r_2-r_1}{4}\left(T_{ia}^+T_{jb}^+
+ T_{ib}^-T_{ja}^- \right. \nonumber\\
& &\left. \hskip 3cm + T_{ib}^+T_{ja}^+ + T_{ia}^-T_{jb}^-
\right)\nonumber\\
&+&\! J\alpha\left\{\frac{r_2-r_1}{4}\left[n_{ic}\left(1-n_{jc}\right)
+\left(1-n_{ic}\right)n_{jc}\right]\right. \nonumber\\
& &\!\left. \hskip .5cm +\frac{2r_2+r_3}{3}\;n_{ic}n_{jc}\right\}\,.
\end{eqnarray}
The operators for the bonds along two other lattice directions can be
obtained by permutations of $\{a,b,c\}$ orbital indices. However, as
all the bonds are equivalent, it suffices to consider the above
operator ${\hat {\cal J}}_{ij}^{(c)}$ for a representative bond to
derive the exchange constant by averaging the orbital operators over
the GS wave function $|\Phi_0\rangle$,
\begin{equation}
\label{jmf}
J_{\rm MF}\equiv
\langle\Phi_0|{\hat{\cal J}}_{ij}^{(\gamma)}|\Phi_0\rangle\,.
\end{equation}
Here the orbital fluctuation operators in the term
$\propto\sqrt{\alpha(1-\alpha)}$ contribute and couple different
components of the wave function $|\Phi_0\rangle$.

We also consider a simplified classical quantity,
\begin{equation}
\label{jmf0}
J_{\rm MF}^0\equiv \sum_n
\langle n|{\hat{\cal J}}_{ij}^{(\gamma)}|n\rangle\,
|\langle n|\Phi_0\rangle|^2\,,
\end{equation}
where the summation includes basis states $\{|n\rangle\}$ with all
possible orbital configurations in the considered cluster. As the
basis states are used for calculating the average in Eq. (\ref{jmf0})
which is the same along all the bonds. This result may be derived from
Eq. (\ref{jorbi}) by neglecting the orbital dynamics, {\it inter alia\/}
the terms $\propto\sqrt{\alpha(1-\alpha)}$, and keeping only
the diagonal orbital terms. One finds then the approximate form of
the spin interaction,
\begin{eqnarray}
\label{jorbi0}
{\bar {\cal J}}_{ij}^{(c)}\!\! &=&\!
J(1-\alpha)\left\{
-\frac{3r_1+r_2+2r_3}{3}\left(
T_{ic}^z T_{jc}^z - \frac14 n_i^{(c)}n_j^{(c)}\right)\right.
\nonumber\\
& &\hskip 1.4cm +\left.\frac{r_1-r_2}{4}
\left(n_{ic}+n_{jc}-2\right)\right\}
\nonumber\\
&+&\!J\alpha\left\{\frac{r_2-r_1}{4}
\left[n_{ic}\left(1-n_{jc}\right)
+\left(1-n_{ic}\right)n_{jc}\right]\right. \nonumber\\
& &\!\left. \hskip .5cm +\frac{2r_2+r_3}{3}\;n_{ic}n_{jc}\right\}
\,.
\end{eqnarray}
Further simplification follows from an observation that for uniform
electron distribution in the spin-orbital liquid state one finds the
following averages for the relevant density and pseudospin operators:
\begin{equation}
\label{avesol}
\left\langle n_{i\gamma}\right\rangle=\frac{1}{3},\hskip .5cm
\left\langle n_i^{(\gamma)}\right\rangle=\frac{2}{3},\hskip .5cm
\left\langle T_{i\gamma}^z T_{j\gamma}^z\right\rangle=0\,.
\end{equation}
Using these expectation values in Eq. (\ref{jmf0}) it follows,
\begin{eqnarray}
\label{jorbi0av}
J_{\rm MF}^0&=& \frac{1}{3}\,J(1-\alpha)\left\{
\frac{3r_1+r_2+2r_3}{9}-r_1+r_2\right\}
\nonumber\\
&+&\frac{1}{9}\,J\alpha\left\{r_2-r_1+\frac{2r_2+r_3}{3}\right\}
\,,
\end{eqnarray}
and one arrives at an analytic expression,
\begin{equation}
\label{jmf0av}
J_{\rm MF}^0=J\,(2-\alpha)\;\frac{-3r_1+5r_2+r_3}{27}\,.
\end{equation}
The classical exchange constant $J_{\rm MF}^0$ Eq. (\ref{jmf0av}) is AF
for small values of $\eta$, in agreement with the results presented in
Secs. \ref{sec:tri} and \ref{sec:pbc}, and changes sign at the critical
value of Hund's exchange,
\begin{equation}
\label{etac}
\eta_c=0.188\,,
\end{equation}
where the present classical evaluation of the exchange constant
predicts a transition to the FM phase. The present classical treatment
suggests that this transition would occur simultaneously for all the
bonds directly from the low-spin to high-spin state, with the maximal
value of total spin ${\cal S}=N/2$ for the cluster of $N$ sites.

\begin{figure}[t!]
\includegraphics[width=8.2cm]{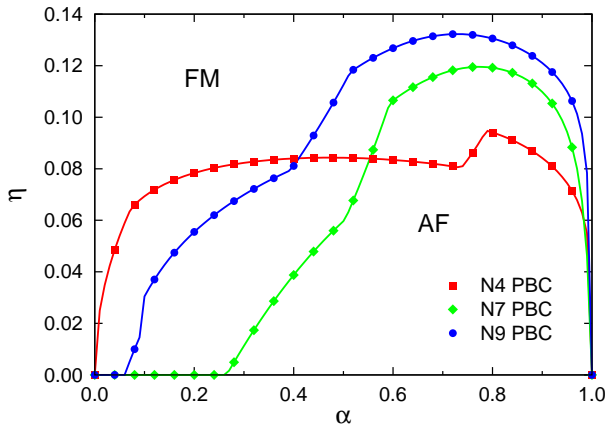}
\caption{(Color online) Phase diagrams obtained for disentangled
spin-orbital interactions following Eq. (\ref{sommf}) obtained for N4,
N7 and N9 clusters with PBC. No intermediate phases were found between
low-spin (${\cal S}={\cal S}_{\rm min}$) and high-spin
(${\cal S}={\cal S}_{\rm min}$) phases.}
\label{fig:mf}
\end{figure}

As usually, the MF Hamiltonian Eq. (\ref{sommf}) implies a
self-consistent solution of spin and orbital correlations. For
instance, by solving a similar 1D problem within the MF
approach self-consistently, one finds dimerization in FM spin-orbital
chains at finite temperature.\cite{Sir08} Here we have applied the
following iterative procedure for a system with spin and orbital
interactions assumed to be isotropic on the triangular lattice:
(i) for an initial value of spin scalar product
$\left\langle{\vec S}_i\cdot {\vec S}_j\right\rangle$ we solved the
orbital Hamiltonian, next
(ii) the effective exchange constants given by Eq. (\ref{jmf}) were
obtained, and
(iii) they were used to determine the spin scalar product. This cycle
was repeated until a self-consistent solution was found. We also used
certain damping along the iteration process and enforced the symmetry
of the considered clusters to accelerate the convergence as all the
bonds are equivalent when PBC are used.

The phase diagrams obtained following the above MF procedure for the
clusters N4, N7 and N9 with PBC are presented in Fig. \ref{fig:mf}. The
applied procedure does not use the total spin symmetry, so intermediate
(partial) spin polarization of the cluster could not be resolved.
Since we solve here the effective spin Hamiltonian independently of the
orbital problem, the
total spin state is determined entirely by the sign of the exchange
constant $J_{\rm MF}$. The spin correlations were found to be either
negative when $J_{\rm MF}>0$, indicating local AF (singlet-like)
correlations, or were classical and FM, i.e.,
$\left\langle{\vec S}_i\cdot
{\vec S}_j\right\rangle=+\frac14$ when $J_{\rm MF}<0$. One
finds that the low-spin states are stable in a narrower range of the
$(\alpha,\eta)$ phase diagram for all three considered clusters, see
Fig. \ref{fig:mf}, than when the exact diagonalization of the full
Hamiltonian (\ref{som}) is performed (shown in Fig. \ref{fig:phdpbc}).
In general, the phase boundary between the low-spin
(${\cal S}={\cal S}_{\rm min}$) and high-spin
(${\cal S}={\cal S}_{\rm max}$) phase
was found at a lower value of $\eta$ in each considered cluster than
for the data extracted from exact diagonalization.

Moreover, the superexchange and direct exchange cases are rather
special and FM states appear in these limits already for infinitesimal
values of $\eta$. While the low-spin and high-spin
state are degenerate in the direct exchange limit (at $\alpha=1$) for
all three clusters shown in Fig. \ref{fig:mf}, in the superexchange
case (at $\alpha=1$) such a degeneracy was found only for the N4
cluster. Both larger clusters N7 and N9 are FM already at $\eta=0$
in a range of small values of $\alpha$: for $\alpha<0.27$ in case of
N7 and for $\alpha<0.05$ in case of N9 cluster. This behavior is
surprising and suggests that the present MF procedure is unable to
describe the present spin-orbital problem in a realistic way.
We address this question in more detail in the next Sec. \ref{sec:exc}.

\subsection{Effective spin exchange constants}
\label{sec:exc}

In a spin system, as the one obtained from the spin-orbital model,
intersite spin correlations follow the sign of the exchange constant,
i.e., when the exchange constant changes sign and becomes negative,
the spins align in the FM phase. The spin model extracted in the MF
approximation from the spin-orbital model Eq. (\ref{som}) is given by
\begin{equation}
\label{spinmf}
{\cal H}_{s} = J_{\rm MF}
\sum_{\langle ij \rangle}
{\vec S}_i \cdot {\vec S}_j
\,,
\end{equation}
where the exchange constant is given by Eq. (\ref{jmf}).

\begin{figure}[t!]
\includegraphics[width=8cm]{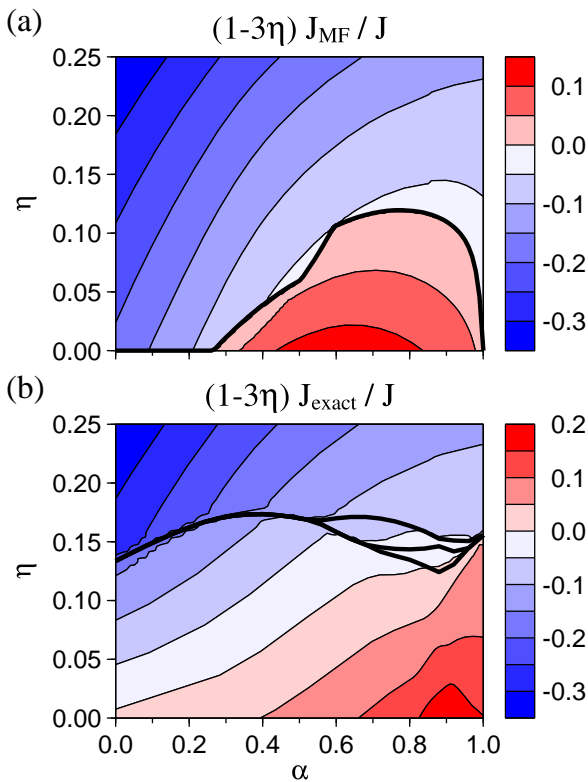}
\caption{(Color online)
Contour plots of the effective exchange constant $J_{\rm MF}$ as
obtained for N7 cluster with PBC from Eq. (\ref{jmf}):
(a) within the MF calculation which includes orbital fluctuations, and
(b) using exact GS found in exact diagonalization.
In case (a) the transition from low-spin to high-spin phase occurs when
the exchange constant $J_{\rm MF}$ changes sign and becomes negative.
Thick lines in (b) indicate the phase boundaries obtained between
phases with increasing total spin value ${\cal S}=1/2$, 3/2, 5/2 and
7/2 for increasing $\eta$.}
\label{fig:jeff}
\end{figure}

Let us consider now once again the N7 cluster as a representative case
to contrast the results obtained from the exact diagonalization and the
present MF approach. The low-spin phase has ${\cal S}=1/2$, and the
transition takes place to the ${\cal S}=7/2$ phase. Knowing that all
the bonds are equivalent when PBC are used, and that each site has six
neighbors, one finds that the spin correlation function for the
low-spin ${\cal S}=1/2$ phase is
$\left\langle\vec{S}_i\cdot\vec{S}_j\right\rangle=-3/28$, as deduced
from Eq. (\ref{ident}). The phase diagram shown in Fig. \ref{fig:mf} may
be understood as following from the change of sign of the MF exchange
constant Eq. (\ref{jmf}). Indeed, when the effective exchange constant
is obtained for the entire $(\alpha,\eta)$ plane, the onset of FM phase
corresponds to the line $J_{\rm MF}=0$, see Fig. \ref{fig:jeff}(a).
In general, the value of $J_{\rm MF}$ decreases with increasing $\eta$
for any value of $\alpha$, but positive values of the effective
exchange constant are found only in a range of $\eta$ values, if
$0.27<\alpha<1$.
This situation is unusual as the effective superexchange obtained in
the MF approach favors FM spin order, even in the absence of Hund's
exchange. It could be understood as a consequence of strong orbital
fluctuations in this regime which provide another mechanism of FM
interactions playing an important role in the magnetic properties
of the $R$VO$_3$ perovskites ($R$=La,Y, {\it etcetera}).\cite{Kha01,Hor08}

We remark that the phase diagram of N7 cluster (see Fig.
\ref{fig:phdpbc}) is quite different from the transition line shown in
Fig. \ref{fig:jeff}(a). Therefore, we conclude that the MF decoupling
scheme does not capture the essential features of the joint spin-orbital
dynamics which stabilizes the low-spin phase in a broad regime of
parameters. Furthermore, the results shown in Fig. \ref{fig:jeff}(b)
prove that spin transition in the spin-orbital model is not related to
the apparent sign change in the exchange constant $J_{\rm MF}$ Eq.
(\ref{jmf}) when it is calculated within
the exact diagonalization approach. This also demonstrates that the
frequently used MF procedure to extract the spin exchange constants
\cite{Ole05} might lead to uncontrolled results, particularly for
frustrated systems with disorder in a form of spin-orbital liquid.

Comparing the values of $J_{\rm MF}$ obtained using the MF procedure and
the exact diagonalization [Figs.~\ref{fig:jeff}(a) and
\ref{fig:jeff}(b), respectively], a qualitative change is found for the
direct exchange case ($\alpha=1$), where positive values of
$J_{\rm MF}$ extend now up to the transition point $\eta\simeq 0.156$.
Also for lower values of $\alpha$ the range of $J_{\rm MF}>0$ is
extended, particularly close to $\alpha=0$. However, the value of
$\eta$ corresponding to $J_{\rm MF}=0$ systematically decreases with
decreasing $\alpha$ and is as low as $\tilde\eta_c\simeq 0.01$ at
$\alpha=0$. This behavior is in drastic contrast with the obtained
transition from the low-spin to high-spin state, which occurs within
the exact diagonalization method at a much higher value of
$\eta\simeq 0.16$ in the entire range of $\alpha$.
Therefore, we conclude that the MF decoupling procedure given in Eq.
(\ref{sommf}) cannot be used in the present situation, similar as for
the entangled states in the 1D spin-orbital chains with active
$t_{2g}$ orbitals.\cite{Ole06}

\section{SUMMARY AND CONCLUSIONS}
\label{sec:summa}

The present study unravels dimer correlations present in the
spin-orbital model on the triangular lattice. By considering finite
clusters with open boundary conditions we have shown that such
dimer correlations do exist when the interactions are either of the
superexchange or of the direct exchange type. When the symmetry of the
lattice is broken by the boundary in triangular clusters, spin singlets
are favored on bonds which originate from corner sites, and the orbital
flavors adjust to the stronger channel, either to superexchange or to
direct exchange. In fact,
as different orbital states support these two types of magnetic
interactions, the orbital distribution over the lattice is partly
frustrated in the entire parameter regime, and this frustration
contributes to the orbital disorder in the crossover regime, typically
around $\alpha=0.6$. The most striking result here is the collapse of
valence-bond states in the intermediate regime and the onset of
spin-orbital liquid. Such a disordered state realized already in finite
clusters with open boundary conditions in the regime of competing
interactions suggests that it could extend over a broader regime of
parameters when the geometry would not favor particular way of symmetry
breaking in the spin-orbital space.

Our study of the clusters with periodic boundary conditions provides
indeed evidence that a quantum spin-orbital liquid phase is realized
in the present $d^1$ spin-orbital model designed for $t_{2g}$ electrons
on the geometrically frustrated triangular lattice. We would like to
emphasize here that the frustrated lattice is necessary to remove
tendency towards certain type of orbital order which could break the
lattice symmetry in other cases, such as in the titanium or vanadium
perovskites, and would support phases with long-range orbital
accompanied by spin order of certain kind (usually following the
Goodenough-Kanamori rules\cite{Goode}). Thus, the spins behave here
differently than in the spin model with Heisenberg interactions on the
triangular lattice, and no spin order emerges when the spins couple to
orbitals and both degrees of freedom undergo joint quantum fluctuations.

Although a mathematical proof is not possible, we provided, as we think,
a rather complete and convincing evidence that the present $d^1$
spin-orbital model realizes a paradigm of {\it spin-orbital liquid
phase\/},
and the order-out-of-disorder mechanism does not apply when the Hilbert
space contains coupled spin and orbital sectors. Previous search for
this quantum state of matter in other systems, particularly in LiNiO$_2$
where $e_g$ orbitals are active on the triangular lattice,\cite{Ver04}
were unsuccessful.\cite{Mos02,Rei05} After considering the present
model in more detail we suggest that the triple degeneracy of $t_{2g}$
orbitals plays a crucial role in the onset of spin-orbital liquid, as
the number of orbital flavors fits to the geometry of the triangular
lattice. However, one might expect that instead a three-sublattice
ordered state could arise, similar to the one known for spin system.
\cite{Faz99}
We argue that the coupling to the spins plays here a very important
role and spin-orbital entanglement is a characteristic feature of the
disordered state found in the absence of Hund's exchange (at $J_H=0$).

We have shown than both geometry and spin-orbital interactions are the
origin of frustration in the model under consideration. One may wonder
in this context whether {\it geometrical frustration\/} on the triangular
lattice enhances {\it interaction frustration\/} for spin-orbital
models. Quite generally, spin-orbital models contain in principle
more channels which can be used for relieving enhanced frustration so
one might still expect that some kind of ordered states would emerge.
We argue that this problem is more subtle and its essence lies in the
nature of spin-orbital entangled states. We have demonstrated by
evaluating spin-orbital correlations that spin and orbital operators
are entangled on the bonds and cannot be factorized. Under these
circumstances important contributions to the ground state energy arise
from joint spin-orbital fluctuations.

We have also found that
Goodenough-Kanamori rules\cite{Goode} are not obeyed by spin and
orbital bond correlations in some cases. This concerns in particular
the superexchange regime where the low-spin phase is stabilized by
them. Such entangled states play an important role in the vanadium
perovskites at finite temperature,\cite{Kha01,Hor08} are lead to
topological constraints on the hole motion which couples simulataneosly
to spin and optical exciatation when states with entangled spin-orbital
order are doped.\cite{Woh09} We emphasize that the spin-orbital
entanglement occurs here on the bonds, and should not be confused with
spin-orbital singlets arising from strong on-site spin-orbit coupling
which leads to spin-orbital liquid with local singlets,\cite{Che09}
and might also generate exotic phases, as shown recently in the case of
spin $S=1/2$ and a higher orbital quantum number (pseudospin) $L=1$.
\cite{Jac09}

By considering the magnetic transition to the ferromagnetic phase, we
have shown that a likely scenario for this transition is a crossover
via the intermediate spin states before fully polarized ferromagnetic
state sets in. This suggests that spin-orbital entangled states also
play an important role in phases with partial spin polarization,
stabilizing them in the regime of transition towards the fully
polarized ferromagnetic phase. Moreover, we detected
a general principle concerning the applicability of effective spin
models derived from spin-orbital Hamiltonians. While this is a common
practice nowadays which helps to understand and interpret the
experimental data in systems with active orbital degrees of freedom,
\cite{Ole05} we presented evidence that even in case when magnetic
exchange constants can be accurately evaluated using the relevant
orbital correlations, they might be inadequate to describe the magnetic
ground state and excitations in such a system. This qualitative
limitation could play a role particularly in disordered systems, where
the physical consequences of entangled spin-orbital states are more
severe.

In summary, the present spin-orbital model on the triangular lattice
provides a beautiful example of highly frustrated interactions with
the ground state dominated by:
(i) quantum fluctuations, and
(ii) spin-orbital entanglement.
It is for this reason that several naive expectations
which have their roots in classical expectations for complex spin
systems do not apply, and it is even impossible to describe correctly
the interactions for the magnetic degrees of freedom by decoupling them
from the orbital ones. Although lattice distortions and coupling
between the planes of a triangular lattice might destabilize the
spin-orbital liquid found here, we hope that its experimental example
could be established by future experimental studies.
The spin-orbital disordered state provides a challenge both for the
theory and for the experiment to find a way of describing magnetic
excitations arising in a spin-orbital liquid phase.

\acknowledgments

We thank George Jackeli, Bruce Normand and Karlo Penc for insightful
discussions.
J.~C.~thanks the Alexander von Humboldt Foundation for the fellowship
during his stay at Max-Planck-Institut f\"ur Festk\"orperforschung,
and acknowledges support by the Ministry of Education of Czech Republic
under Grant No. MSM0021622410.
A.~M.~O. acknowledges support by the Foundation for Polish Science
(FNP) and by the Polish Ministry of Science and Higher Education
under Project No. N202~069639.

\end{document}